\newcommand{\idty}{{\leavevmode{\rm 1\mkern -5.4mu I}}}
\newcommand{\Ibb}[1]{ {\rm I\ifmmode\mkern
            -3.6mu\else\kern -.2em\fi#1}}
\newcommand{\ibb}[1]{\leavevmode\hbox{\kern.3em\vrule
     height 1.2ex depth -.3ex width .2pt\kern-.3em\rm#1}}
\newcommand{\Nl}{{\Ibb N}}
\newcommand{\Cx}{{\ibb C}}
\newcommand{\Rl}{{\Ibb R}}
\newcommand{\be}{\begin{eqnarray}}
\newcommand{\ee}{\end{eqnarray}}

\documentstyle[12pt]{article}
\begin{document}

\renewcommand{\theequation} {\arabic{section}.\arabic{equation}}

\begin{tabbing}
\hspace*{12cm}\= MPI-PhT/94-83 \\
              \> hep-th/9412069 \\
              \> December 1994
\end{tabbing}
\vskip.5cm

\centerline{\Large \bf Differential Calculi on Commutative Algebras}
\vskip1cm

\begin{center}
  {\bf H.C. Baehr$^\natural$, A. Dimakis$^\sharp$} and
  {\bf F. M\"uller-Hoissen$^\natural {}^\star$}
      \vskip.5cm
     $^\natural$Institut f\"ur Theoretische Physik,
      Bunsenstr. 9, D-37073 G\"ottingen
      \vskip.2cm
     $^\sharp$Department of Mathematics, University of Crete,
      GR-71409 Iraklion
        \vskip.2cm
     $^\star$Max-Planck-Institut f\"ur Physik,
     Werner-Heisenberg-Institut \\
     F\"ohringer Ring 6, D-80805 Munich$^\dagger$

\end{center}

\vskip1.cm

\begin{abstract}
\noindent
A differential calculus on an associative algebra $\cal A$ is an
algebraic analogue of the calculus of differential forms on a
smooth manifold. It supplies $\cal A$ with a structure on which
dynamics and field theory can be formulated to some extent in
very much the same way we are used to from the geometrical arena
underlying classical physical theories and models.
In previous work, certain differential calculi on a commutative
algebra exhibited relations with lattice structures, stochastics,
and parametrized quantum theories. This motivated the present
systematic investigation of differential calculi on commutative
and associative algebras. Various results about their structure
are obtained. In particular, it is shown that there is a correspondence
between first order differential calculi on such an algebra and
commutative and associative products in the space of 1-forms.
An example of such a product is provided by the It{\^o} calculus
of stochastic differentials.
For the case where the algebra $\cal A$ is freely generated by
`coordinates' $x^i, \, i=1,\ldots,n,$ we study calculi for which the
differentials $\mbox{d}x^i$ constitute a basis of the space of
1-forms (as a left $\cal A$-module).
These may be regarded as `deformations' of the ordinary
differential calculus on $\Rl^n$. For $n \leq 3$ a classification of
all (orbits under the general linear group of) such calculi with
`constant structure functions' is presented. We analyse whether these
calculi are reducible (i.e., a skew tensor product of
lower-dimensional calculi) or whether they are the extension (as defined
in this article) of a one dimension lower calculus. Furthermore,
generalizations to arbitrary $n$ are obtained for all these calculi.
\end{abstract}

\vskip1cm
$^\dagger$Temporary address.

\newpage

\renewcommand{\contentsname}{\normalsize \bf Contents}
\small
\tableofcontents
\normalsize

\newpage

\section{Introduction}
\setcounter{equation}{0}
During the last years there has been a rapid increase of interest
in `noncommutative geometry'. Basically, this notion stands for an
attempt to get away from the classical concept of a (differentiable)
manifold as the arena on which physics takes place. In particular,
this is strongly motivated by considerations about space-time
structure at very small length scales, and quantum gravity.
The manifold is replaced by some abstract algebra $\cal A$ which is
usually assumed to be associative, but not necessarily commutative.
In order to be able to formulate dynamics and field theories on or
with such `generalized spaces', a convenient tool appears to be a
`differential calculus' on it which is an algebraic analogue of the
calculus of differential forms on a manifold.\footnote{Such a point of
view has been pioneered by Robert Hermann \cite{Herm77}.}
\vskip.2cm

If the algebra $\cal A$ is commutative, then one can construct a
(topological) space on which it can be realized as an algebra of
functions. Besides the familiar continua this case also includes finite
or, more generally, discrete spaces.
Differential calculi on commutative algebras have been considered
and explored in several papers (see \cite{Conn+Lott90}-\cite{DMH94-discr},
for example). If the algebra $\cal A$ is (freely) generated by
`coordinates' $x^k, \, k=1, \ldots,n$ (together with a unit), a
differential calculus on it can be specified via commutation relations
with their `differentials',
\be                         \label{dc-intro}
   \lbrack \mbox{d}x^k , x^\ell \rbrack = C^{k\ell}{}_m \,
                                          \mbox{d}x^m
\ee
where $C^{k\ell}{}_m \in {\cal A}$ (subject to certain
constraints).\footnote{On the rhs of the last equation and in the
following we are using the summation convention if not
stated otherwise.}
An example of interest for physics is given by
\be
   \lbrack \mbox{d}x^k , x^\ell \rbrack = a^k \, \delta^{k\ell} \,
           \mbox{d}x^k  \qquad (\mbox{no summation})
\ee
which may be regarded as the basic structure underlying lattice
theories \cite{DMHS93} ($a^k$ plays the role of the lattice
spacing in the $k$th direction). Another example is
\be                 \label{ex-grav}
   \lbrack \mbox{d}x^k , x^\ell \rbrack = \gamma \, g^{k\ell} \,
                                       \mbox{d}x^{n+1} , \quad
   \lbrack \mbox{d}x^{n+1} , x^k \rbrack = 0
\ee
where $g^{k \ell}$ are the components (with respect to coordinates
$x^i$ on a manifold) of a real contravariant tensor field. For
$\gamma = i \, \hbar$ this may be viewed as a basic structure
underlying parametrized (proper time) quantum theories
\cite{DMH92-grav}. For real and positive definite $\gamma \, g^{ij}$
one recovers the It{\^o} calculus of stochastic differentials
\cite{DMH93-stoch}. These examples motivate a systematic
investigation of the possibilities. In \cite{DMHS93} all differential
calculi subject to (\ref{dc-intro}) with $n=2$ and constant structure
functions, i.e., $C^{k \ell}{}_m \in \Cx$, were classified.\footnote{In
\cite{BKO93} the case $n=2$ and $C^{k \ell}{}_m$ linear in $x^\ell$ has
been treated. Such differential calculi are also obtained from calculi
on the Heisenberg algebra \cite{DMH92} in the limit $\hbar \to 0$.}
The procedure used there does not extend to $n>2$, however. Here, we
therefore propose an alternative method and present the classification
of 3-dimensional calculi (see also \cite{Baeh94}).
\vskip.2cm

Section 2 recalls some basic definitions and constructions used in the
sequel. Section 3 presents general results about differential
calculi on a commutative (and associative) algebra $\cal A$. In
particular, it is shown that every (first order) differential calculus
on $\cal A$ determines an $\cal A$-bilinear
commutative and associative product in the space of 1-forms.
This relates the problem of classifying (first order)
differential calculi to that of determining all
$\cal A$-bimodules over $\cal A$ with such product structures.
This correspondence generalizes the relation established in
\cite{DMH93-stoch} between the It{\^o} calculus of stochastic
differentials (where one has a product in the space of 1-forms)
and a differential calculus of the form (\ref{ex-grav}).
\vskip.2cm

In section 4 we consider the case where $\cal A$ is freely generated
(as a commutative and associative algebra) by elements
$x^i, \, i=1,\ldots,n,$ together with a unit $\idty$. The class
of differential calculi for which the set of differentials $\mbox{d}x^i$
are a basis of the space of 1-forms (as a left $\cal A$-module) is then
explored in some detail. They may be regarded as deformations of the
ordinary differential calculus on $\Rl^n$ and are therefore of
special interest. We then address the classification
problem for such calculi with constant structure functions and describe
corresponding results. The action of the `exterior derivative'
$\mbox{d}$ determines left- and right-partial derivatives
$D_{\pm i} \, : \, {\cal A} \rightarrow {\cal A}$ via
\be
      \mbox{d} f = (D_i f) \, \mbox{d} x^i
                 = \mbox{d} x^i \, D_{-i} f
                   \qquad (\forall f \in {\cal A}) \; .
\ee
They display the most important properties of a differential calculus.
Some general results concerning their structure are obtained (see
also section 3.3).
Examples are provided by the irreducible calculi which arose from our
classification of $n=3$ calculi.
\vskip.2cm

Some of our results extend to a certain generalization of the notion of
a differential calculus and this is the subject of an appendix.
Section 5 contains our conclusions.

\section{Algebraic differential calculi on associative algebras}
\setcounter{equation}{0}
In this section some basic algebraic constructions are recalled
which are needed in the following sections.
\vskip.3cm

Let $\cal A$ be an associative algebra over\footnote{Most of the
following also works over $\Rl$ (or other fields), but for the
classification results in section 4 the choice $\Cx$ is essential.}
$\Cx$ with unit $\idty$. A {\em differential calculus}\footnote{In the
mathematical literature it is usually called a {\em differential graded
algebra}.} $(\Omega({\cal A}),\mbox{d})$ over $\cal A$ is a graded
associative algebra
\be
   \Omega({\cal A}) = \bigoplus_{r=0}^\infty \Omega^r({\cal A})
\ee
where $\Omega^r({\cal A})$ are $\cal A$-bimodules and
$\Omega^0({\cal A}) = \cal A$. It is supplied with a linear operator
of degree 1
\be
 \mbox{d} \, : \, \Omega^r({\cal A}) \rightarrow \Omega^{r+1}({\cal A})
\ee
satisfying $ \mbox{d}^2=0$, $\mbox{d} \idty = 0$ and
\be                    \label{Leibniz}
  \mbox{d}(\omega \omega') = (\mbox{d}\omega) \, \omega'
    + (-1)^r \omega \, \mbox{d} \omega'
\ee
where $\omega \in \Omega^r({\cal A})$. d is called
{\em exterior derivative}. We also demand that, for
$r>0$, $\Omega^r({\cal A})$ is {\em generated} by $\mbox{d}$ in the
sense that $\mbox{d} \Omega^{r-1}({\cal A})$ generates
$\Omega^r({\cal A})$ as an $\cal A$-bimodule. This additional
assumption will be relaxed in the appendix. We also assume that
$\Omega({\cal A})$ is unital with unit $(\idty,0,\dots)$. The
elements of $\Omega^r({\cal A})$ are called {\em r-forms}.
\vskip.3cm

$(\Omega^1({\cal A}),\mbox{d})$ (with $\mbox{d}$ restricted to $\cal A$)
is a {\em first order differential calculus} on $\cal A$. $\mbox{d}$ is
then a derivation ${\cal A} \rightarrow \Omega^1({\cal A})$.

\subsection{The universal first order differential calculus}
The tensor product ${\cal A} \otimes {\cal A}$ consists of finite
linear combinations (with coefficients in $\Cx$) of terms $f \otimes
h$ where $f,h \in {\cal A}$. With the multiplication
\be            \label{tensorprod}
  (f \otimes h) \, (f' \otimes h') := ff' \otimes h h'
  \qquad \forall \, f,f',h,h' \in {\cal A}
\ee
it becomes an associative algebra (over $\Cx$). Via
\be
  g \, (f \otimes h) &:=& (g \otimes \idty) (f \otimes h)
  = (g f) \otimes h       \nonumber \\
  (f \otimes h) \, g &:=& (f \otimes h) (\idty \otimes g)
  = f \otimes (h g)                \label{univ-bimod}
\ee
${\cal A} \otimes {\cal A}$ carries an $\cal A$-bimodule structure.
The multiplication in $\cal A$ yields a map
\be
  \mu \, : \, {\cal A} \otimes {\cal A} \rightarrow {\cal A}
  \, , \quad f \otimes h \mapsto f h
\ee
which is a bimodule homomorphism (but not an algebra homomorphism,
in general). Let us define
\be
   \tilde{\Omega}^1({\cal A}) := \mbox{ker } \mu
   = \lbrace \sum_a f_a \otimes h_a \, \mid \, \sum_a f_a h_a = 0
     \rbrace  \; .
\ee
Then we have a map
\be                         \label{univ-d}
 \tilde{\mbox{d}} \, : \, {\cal A} \rightarrow
 \tilde{\Omega}^1 ({\cal A}) \; , \quad
 f \mapsto \idty \otimes f - f \otimes \idty  \; .
\ee
The image of $\cal A$ under $\tilde{\mbox{d}}$ generates
$\tilde{\Omega}^1({\cal A})$ as an $\cal A$-bimodule.
$(\tilde{\Omega}^1({\cal A}), \tilde{\mbox{d}})$ is the {\em universal
first order differential calculus} on $\cal A$. It
has the following universal property.
\vskip.3cm
\noindent
{\bf Theorem 2.1} For each derivation $\mbox{d} \, : \, {\cal A}
\rightarrow M$ into some $\cal A$-bimodule $M$ there is one and
only one $\cal A$-bimodule homomorphism $\phi \, : \,
\tilde{\Omega}^1({\cal A}) \rightarrow M$ such that $\mbox{d}
= \phi \circ \tilde{\mbox{d}}$, i.e., the following diagram
commutes,

\unitlength=1.mm
\linethickness{0.4pt}
\begin{picture}(80.,24.)
\put(66.00,3.00){\makebox(0,0)[cc]{$M$}}
\put(56.00,17.00){\makebox(0,0)[cc]{$\cal A$}}
\put(82.00,17.00){\makebox(0,0)[cc]{$\tilde{\Omega}^1({\cal A})$}}
\put(66.00,20.00){\makebox(0,0)[cc]{$\tilde{\mbox{d}}$}}
\put(74.00,9.00){\makebox(0,0)[cc]{$\phi$}}
\put(57.00,9.00){\makebox(0,0)[cc]{$\mbox{d}$}}
\put(75.00,14.00){\vector(-3,-4){6.67}}
\put(57.00,14.00){\vector(3,-4){6.67}}
\put(59.00,17.00){\vector(1,0){15.00}}
\end{picture}

\vskip.1cm
\noindent
{\bf Proof:} see \cite{Bour89}, chapter III, \S 10.10, for example.
                                           \hfill  {\Large $\Box$}
\vskip.3cm
\noindent
As a consequence of this theorem, every first order differential
calculus $(\Omega^1({\cal A}),\mbox{d})$ on $\cal A$ (which is
generated by $\mbox{d}$) is isomorphic to a quotient of
$\tilde{\Omega}^1({\cal A})$ by some $\cal A$-subbimodule (the kernel
of the respective homomorphism $\phi$).

\subsection{The universal differential calculus}
Let $(\tilde{\Omega}^1({\cal A}), \tilde{\mbox{d}})$ be the universal
first order differential calculus. Define
\be
  \tilde{\Omega}^0({\cal A}) := {\cal A} , \quad
  \tilde{\Omega}^p({\cal A}) := \underbrace{\tilde{\Omega}^1({\cal A})
  \otimes_{\cal A} \cdots \otimes_{\cal A} \tilde{\Omega}^1({\cal A})}
  _{\mbox{$p$ times}}  \; .
\ee
Then
\be
   \tilde{\Omega}({\cal A}) := \bigoplus^\infty_{p=0} \tilde{\Omega}^p
   ({\cal A})
\ee
with the multiplication $\otimes_{\cal A}$ becomes a graded
associative algebra. The extension of $\tilde{\mbox{d}}$
to an exterior derivative is given by
\be
  \tilde{\mbox{d}} (f_0 \otimes f_1 \otimes \cdots \otimes f_p)
 := \sum^{p+1}_{q=0} (-1)^q f_0 \otimes \cdots \otimes f_{q-1}
    \otimes \idty \otimes f_q \otimes \cdots \otimes f_p \;
\ee
and \Cx-linearity.
$(\tilde{\Omega}({\cal A}), \tilde{\mbox{d}})$ is the {\em universal
differential calculus}\footnote{It is Karoubi's {\em differential
envelope} of $\cal A$ (see \cite{Kast88}).} on $\cal A$. It has a
universal property generalizing Theorem 2.1 (see \cite{Kast88},
for example).
Any differential calculus on $\cal A$ (for which $\mbox{d}
\Omega^p({\cal A})$ generates $\Omega^{p+1}({\cal A})$ as an $\cal
A$-bimodule) can be obtained from $(\tilde{\Omega}({\cal A}),
\tilde{\mbox{d}})$ as a quotient with respect to some two-sided
differential ideal in $\tilde{\Omega}({\cal A})$ (an ideal which is
closed under $\tilde{\mbox{d}}$).

\subsection{Reducibility and skew tensor products of differential
            calculi}
Let $(\Omega({\cal A}),\mbox{d})$ and $(\Omega({\cal A}'),\mbox{d}')$
be differential calculi over ${\cal A}$ and ${\cal A}'$,
respectively. From these one can build the differential calculus
$(\Omega({\cal A}) \hat{\otimes} \Omega({\cal
A}'),\hat{\mbox{d}})$, called the {\em skew tensor product} (cf
\cite{Kast88}, Appendix A, and \cite{Greu67}, chapter II
\footnote{Instead of {\em skew} the term {\em anticommutative} is used
there.}). The underlying set is the tensor product $\Omega({\cal A})
\otimes \Omega({\cal A}') =: \hat{\Omega}$. The grading is given
by
\be
  \hat{\Omega} = \bigoplus^\infty_{r=0} \hat{\Omega}^r
  \quad \mbox{with} \quad
  \hat{\Omega}^r = \bigoplus^r_{p=0} \Omega^p({\cal A}) \otimes
  \Omega^{r-p}({\cal A}')  \; .
\ee
Multiplication is defined by
\be
  (\omega \hat{\otimes} \omega') (\rho \hat{\otimes} \rho')
       := (-1)^{\partial \omega' \cdot \partial \rho} \,
         ( \omega \rho \hat{\otimes} \omega' \rho' )
\ee
and $\Cx$-linearity. We use $\hat{\otimes}$ to stress the difference
to the canonical multiplication. The linear operator $\hat{\mbox{d}}$
on $\Omega({\cal A}) \hat{\otimes} \Omega({\cal A}')$ acts as follows,
\be
  \hat{\mbox{d}} (\omega \hat{\otimes} \omega') = (\mbox{d} \omega)
  \hat{\otimes} \omega' + (-1)^{\partial \omega} \, \omega
  \hat{\otimes} \mbox{d}' \omega'
\ee
where $\partial \omega$ denotes the grade of the form $\omega$.
\vskip.3cm

Given a differential calculus on an algebra $\cal A$, the question
arises, whether it is {\em reducible} in the sense that it
is a skew tensor product of differential calculi. Otherwise
we should call the differential calculus {\em irreducible}.

\subsection{Inner extensions of derivations}
A derivation $\mbox{d} \, : \, {\cal A} \rightarrow M$ is called
{\em inner} if there is an element $\rho \in M$ such that
\be                  \label{df-inner-rho}
 \mbox{d} f = \lbrack \rho , f \rbrack \qquad \forall f \in {\cal A}
              \; .
\ee
We say that a (first order) differential calculus is inner if its
exterior derivative $\mbox{d}$ is inner.
\vskip.3cm

Given a derivation $\mbox{d} \, : \, {\cal A} \rightarrow M$ (which
may already be inner), the $\cal A$-bimodule $M$ can always be extended
into a larger $\cal A$-bimodule $\check{M}$ such that $\mbox{d}$ becomes
inner. This is done by adding an (independent) element $\rho$ as
follows. Let ${}_{\cal A}\rho$ be the free left $\cal A$-module
generated by $\rho$ and define $\check{M} := M \oplus {}_{\cal A}\rho$
which is then also a left $\cal A$-module. A right $\cal A$-module
structure can then be introduced on $\check{M}$ by requiring
$M \subset \check{M}$ to be an $\cal A$-subbimodule and setting
\be
   (h \, \rho) \, f := h \, f \, \rho + h \; \mbox{d}f
                       \qquad \forall f,h \in {\cal A}  \; .
\ee
Then
\be
   (h \, \rho) \, f = h \, (f \, \rho + \mbox{d}f) = h \, (\rho \, f)
\ee
and
\be
 \rho \, (h f) &=& h \, f \, \rho + \mbox{d}(hf)
                = h \, f \, \rho + h \, \mbox{d}f + (\mbox{d}h) \, f
                = h \, (\rho \, f) + (\mbox{d}h) \, f
                = (h \, \rho + \mbox{d}h) \, f  \nonumber \\
               &=& (\rho \, h) \, f
\ee
which extends the $\cal A$-bimodule structure of $M$ to $\check{M}$.
\vskip.3cm

In some cases it is possible to enlarge the algebra $\cal A$ (by
introducing an additional generator) to an algebra $\check{\cal A}$ and to
extend $\mbox{d}$ such that it becomes inner with an element $\rho$ of
the $\check{\cal A}$-bimodule generated by $\mbox{d} \check{\cal A}$. See
section 4.1.

\section{Differential calculi on commutative algebras}
\setcounter{equation}{0}
In this section, $\cal A$ always denotes an associative and
{\em commutative} algebra. For any first order differential calculus
$(\Omega^1({\cal A}),\mbox{d})$ we have
\be
    \lbrack \mbox{d}f , h \rbrack = \lbrack \mbox{d}h , f \rbrack
\ee
which shows that this commutator is actually a function of $\mbox{d}f$
and $\mbox{d}h$ (and does not depend directly on $f$ and $h$):
\be             \label{C-def}
    \lbrack \mbox{d}f , h \rbrack =: C(\mbox{d}f,\mbox{d}h)   \; .
\ee
$C$ is obviously a bilinear map $\mbox{d}{\cal A} \times
\mbox{d}{\cal A} \rightarrow \Omega^1({\cal A})$ which is symmetric,
i.e.
\be             \label{C-sym}
          C(\mbox{d}f,\mbox{d}h) = C(\mbox{d}h,\mbox{d}f) \, ,
\ee
as a consequence of our first equation. In the following,
it will be shown that $C$ determines an $\cal A$-bilinear associative
and commutative product in the space of 1-forms.

\subsection{The canonical product in the space of universal 1-forms}
For an associative and commutative algebra $\cal A$ also ${\cal A}
\otimes {\cal A}$ with the canonical multiplication rule
(\ref{tensorprod})
is an associative commutative algebra. The map $\mu$ introduced in
section 2.1 then becomes an algebra homomorphism so that $\tilde
{\Omega}^1({\cal A}) = \mbox{ker} \, \mu$ is an ideal.
As a consequence, in the space $\tilde {\Omega}^1({\cal A})$ of 1-forms
of the universal first order differential calculus on $\cal A$ there
is a canonical associative and commutative product
\be
    \tilde {\Omega}^1({\cal A}) \times \tilde {\Omega}^1({\cal A})
    \rightarrow \tilde {\Omega}^1({\cal A}) \, ,  \quad
    (\tilde{\omega} , \tilde{\omega}') \mapsto
    \tilde{\omega} \bullet \tilde{\omega}'
\ee
which is $\cal A$-bilinear, i.e.,
\be
      (f \tilde{\omega} h) \bullet (f' \tilde{\omega}' h')
   = f f' \, (\tilde{\omega} \bullet \tilde{\omega}') \, h h'
\ee
(cf (\ref{univ-bimod})). From
\be                     \label{tens-prod-comm}
  \lbrack \, \sum_a g_a \otimes h_a \, , \, f \, \rbrack
   = \left( \sum_a g_a \otimes h_a \right) (\idty \otimes f
     - f \otimes \idty)
\ee
and (\ref{univ-d}) we deduce the following important property,
\be
    \lbrack \tilde{\omega} , f \rbrack = \tilde{\omega} \bullet
                                         \tilde{\mbox{d}} f
    \qquad \forall \tilde{\omega} \in \tilde{\Omega}^1({\cal A}),
    \, f \in  {\cal A}  \; .
\ee
A simple calculation now leads to
\be                                \label{bull-Leibniz}
 \tilde{\mbox{d}}(fh) = f \, \tilde{\mbox{d}}h + h \, \tilde{\mbox{d}} f
       + \tilde{\mbox{d}} f \bullet \tilde{\mbox{d}} h
\ee
which is generalized in the following Lemma.
\vskip.3cm
\noindent
{\bf Lemma 3.1}
\be
    \tilde{\mbox{d}}(f_1 \cdots f_r) = \tilde{\mbox{d}}
    f_1 \bullet \cdots \bullet \tilde{\mbox{d}} f_r +
    \sum_{k=2}^r {1\over (k-1)!
    \, (r-k+1)!} \, f_{(1} \cdots f_{k-1} \, \tilde{\mbox{d}}
    f_k \bullet \cdots \bullet \tilde{\mbox{d}} f_{r)}    \;
\ee
where the indices on the rhs are totally symmetrised (indicated by
brackets).
\vskip.1cm
\noindent
{\bf Proof:} by induction, using the identity
\begin{eqnarray*}
 f_{(1} \cdots f_{k-1} \, \tilde{\mbox{d}}
    f_k \bullet \cdots \bullet \tilde{\mbox{d}} f_{r+1)}
 &=& {(k-1)! \over (k-2)!} \, f_{r+1} \, f_{(1} \cdots f_{k-2} \,
 \tilde{\mbox{d}} f_{k-1} \bullet \cdots \bullet \tilde{\mbox{d}}
 f_{r)}         \\
 & & + {(r+1-k+1)! \over (r-k+1)!} \,
 f_{(1} \cdots f_{k-1} \, \tilde{\mbox{d}}
 f_k \bullet \cdots \bullet \tilde{\mbox{d}} f_{r)} \bullet
 \tilde{\mbox{d}} f_{r+1}  \; .
\end{eqnarray*}
                               \hfill {\Large $\Box$}
\vskip.3cm
\noindent
A result about 2-forms is expressed next.
\vskip.3cm
\noindent
{\bf Lemma 3.2}
\be
   \tilde{\mbox{d}}(\tilde{\mbox{d}} f_1 \bullet \cdots \bullet
   \tilde{\mbox{d}} f_r)  = - \sum_{k=1}^{r-1} {1 \over k! \,
   (r-k)!} \,
   (\tilde{\mbox{d}} f_{(1} \bullet \cdots \bullet \tilde{\mbox{d}}
   f_k) \, (\tilde{\mbox{d}} f_{k+1} \bullet \cdots \bullet
   \tilde{\mbox{d}} f_{r)})  \; .                 \label{dd-prod}
\ee
\vskip.1cm
\noindent
{\bf Proof:} by induction. In order to show that (\ref{dd-prod})
implies the corresponding formula with $r$ replaced by $r+1$, one may
start with the rhs of the latter and write
\begin{eqnarray*}
  & & (\tilde{\mbox{d}} f_{(1} \bullet \cdots \bullet \tilde{\mbox{d}}
   f_k) \, (\tilde{\mbox{d}} f_{k+1} \bullet \cdots \bullet
   \tilde{\mbox{d}} f_{r+1)})      \\
  &=& k! \, (r+1-k)! \, \sum_{partitions} (\tilde{\mbox{d}} f_{\ell_1}
    \bullet \cdots \bullet \tilde{\mbox{d}} f_{\ell_k}) \,
    (\tilde{\mbox{d}} f_{\ell_{k+1}} \bullet \cdots \bullet
    \tilde{\mbox{d}} f_{\ell_{r+1}})
\end{eqnarray*}
where the sum is taken over all partitions of $(1,\ldots,r+1)$ into
ordered tuples $(\ell_1,\ldots,\ell_k)$, $(\ell_{k+1},\ldots,
\ell_{r+1})$. This sum splits into a sum over partitions with
$r+1 \in \lbrace \ell_1, \ldots, \ell_k \rbrace$ and a sum over
partitions with $r+1 \in \lbrace \ell_{k+1}, \ldots, \ell_{r+1}
\rbrace$. The first of these sums can then be expressed as a sum over
all partitions of $(1,\ldots,r)$ into ordered tuples $(\ell_1,\ldots,
\ell_{k-1})$, $(\ell_k,\ldots,\ell_r)$. The second sum is treated
similarly. The further procedure is then quite evident.
                                             \hfill {\Large $\Box$}

\subsection{The ring structure of a first order differential calculus}
Let $(\Omega^1({\cal A}), \mbox{d})$ be a first order differential
calculus on $\cal A$. In the following, we show that the canonical
product in the space $\tilde{\Omega}^1({\cal A})$ of universal 1-forms
induces a corresponding product in $\Omega^1({\cal A})$.
\vskip.3cm
\noindent
{\bf Lemma 3.3} If $\phi \, : \, \tilde{\Omega}^1({\cal A})
\rightarrow \Omega^1({\cal A})$ is an $\cal A$-bimodule homomorphism,
then $\mbox{ker} \, \phi$ is an ideal in $\tilde{\Omega}^1({\cal A})$
as a ring with product $\bullet$.
\vskip.1cm
\noindent
{\bf Proof:}
An arbitrary element of $\tilde{\Omega}^1({\cal A})$ can be written as
$\sum_a f_a (\idty \otimes h_a - h_a \otimes \idty)$ with $f_a, h_a \in
{\cal A}$. Let $\tilde{\omega} \in \mbox{ker} \, \phi$. Then
\begin{eqnarray*}
  \phi( \tilde{\omega} \bullet \sum_a f_a (\idty \otimes h_a - h_a
  \otimes \idty) ) = \sum_a f_a \, \phi(\lbrack \tilde{\omega} , h_a
  \rbrack) = \sum_a f_a \, \lbrack \phi(\tilde{\omega}) , h_a \rbrack
  = 0   \; .
\end{eqnarray*}
Since $\bullet$ is commutative, this shows that $\mbox{ker} \, \phi$ is
an ideal.
                                                 \hfill {\Large $\Box$}
\vskip.3cm
\noindent
As a consequence of this Lemma and Theorem 2.1, we now obtain the
following result.
\vskip.3cm
\noindent
{\bf Theorem 3.1} For every first order differential calculus
$(\Omega^1({\cal A}),\mbox{d})$ there is a unique $\cal A$-bilinear
associative and commutative product $\bullet$ in $\Omega^1({\cal A})$
such that $ \lbrack \omega , f \rbrack = \omega \bullet \mbox{d} f$.
                                             \hfill {\Large $\Box$}
\vskip.3cm
\noindent
The next Lemma gives a characterization of inner differential
calculi.
\vskip.3cm
\noindent
{\bf Lemma 3.4}  The derivation $\mbox{d}$ of a first order
differential calculus is inner if and only if there
is a unit with respect to $\bullet$.
\vskip.1cm
\noindent
{\bf Proof:} The statement is an immediate consequence of the
relation $\omega \bullet \mbox{d}f = \lbrack \omega , f \rbrack$ taking
into account that $\bullet$ is $\cal A$-bilinear.
                                           \hfill {\Large $\Box$}
\vskip.3cm
\noindent
If a first order differential calculus is inner, i.e., $\mbox{d}f
= \lbrack \rho , f \rbrack$ ($\forall f \in {\cal A}$) with an
element $\rho \in \Omega^1({\cal A})$, then $\rho$ is unique. This
follows from Lemma 3.4 together with the fact that the unit of an
algebra is unique. This in turn implies that, if $\mbox{d}$
is inner, the center of the $\cal A$-bimodule $\Omega^1({\cal A})$ is
trivial, i.e., $ \lbrace \zeta \in \Omega^1({\cal A}) \, \mid \,
\lbrack \zeta , f \rbrack = 0 \; \forall f \in {\cal A} \rbrace =
\lbrace 0 \rbrace$.

\vskip.3cm
\noindent
Let ${\cal I}$ be the two-sided differential ideal in
$\tilde{\Omega}({\cal A})$ generated by $\mbox{ker} \, \phi$. Now
\be
   \Omega({\cal A}) := \tilde{\Omega}({\cal A})/ {\cal I}
\ee
together with $\mbox{d} := \pi \circ \tilde{\mbox{d}}$ is a differential
calculus on $\cal A$. Here, $\pi \, : \, \tilde{\Omega}({\cal A})
\rightarrow \Omega({\cal A})$ is the canonical projection.
The ideal $\cal I$ has a decomposition
\be
     {\cal I} = \bigoplus_{r=0}^\infty {\cal I}^r
\ee
where ${\cal I}^0 = \lbrace 0 \rbrace$ and ${\cal I}^1 = \mbox{ker}
\, \phi$, so that
\be
    \Omega({\cal A}) = \bigoplus_{r=0}^\infty \Omega^r({\cal A})
    \quad \mbox{with} \quad
    \Omega^r({\cal A}) = \tilde{\Omega}^r({\cal A})/{\cal I}^r \; .
\ee
\vskip.3cm
\noindent
{\em Example.}
$\tilde{\Omega}^1({\cal A})^2 := \tilde{\Omega}^1({\cal A}) \bullet
\tilde{\Omega}^1({\cal A})$ is an $\cal A$-subbimodule and also a
two-sided ideal in $\tilde{\Omega}^1({\cal A})$. Hence,
\be
  {\cal K}^1({\cal A}) := \tilde{\Omega}^1({\cal A})/
                          \tilde{\Omega}^1({\cal A})^2
\ee
carries an $\cal A$-bimodule structure too and the induced product
is trivial, i.e., the product of any two elements of
${\cal K}^1({\cal A})$ is equal to zero.
Now Theorem 3.1 shows that all elements of ${\cal K}^1({\cal A})$
commute with all elements of $\cal A$. The extension to forms of
higher grade is called {\em K\"ahler differential calculus}
$({\cal K}({\cal A}),\mbox{d}_{\cal K})$ on $\cal A$ \cite{Kunz86}.

Let $\cal A$ be the algebra generated by $x$ with the relation
$x^N = \idty$ for some $N \in \Nl$. Acting on the latter with the
K\"ahler derivation leads to
\be
       x^{N-1} \, \mbox{d}_{\cal K} x = 0
\ee
which implies $\mbox{d}_{\cal K} x = 0$ so that the K\"ahler derivation
is trivial. In the presence of constraints one is therefore led to
consider {\em noncommutative} differential calculi, where differentials
do not commute with elements of $\cal A$ in general, in order to have
a nontrivial $\mbox{d}$. In particular, this is so for differential
calculi on finite sets \cite{DMH94-discr}.
                                               \hfill {\Large $\Box$}

\vskip.3cm
\noindent
{\em Example.} Let $\cal A$ be the commutative and associative algebra
which is freely generated by elements $t$ and $x$ (and a unit $\idty$).
An example of a differential calculus on $\cal A$ which is not inner
is determined by the commutation relations
\be
   \lbrack \mbox{d}x , x \rbrack = \mbox{d}t , \quad
   \lbrack \mbox{d}x , t \rbrack = 0 = \lbrack \mbox{d}t , t \rbrack
\ee
where we assume that $\mbox{d}x, \mbox{d}t$ is a basis of
$\Omega^1({\cal A})$ as a left- (and right) $\cal A$-module.
For the associated product we have
\be
     \mbox{d}x \bullet \mbox{d}x = \mbox{d}t , \quad
     \mbox{d}x \bullet \mbox{d}t = 0         , \quad
     \mbox{d}t \bullet \mbox{d}t = 0           \; .
\ee
This product is then consistent with commutativity of differentials and
elements of $\cal A$. A realization of this algebra is given by
a stochastic time variable $t$ and a Wiener process $x=W_t$. The
above relations are basic formulas in the It{\^o} calculus of
stochastic differentials (where $\mbox{d}$ is not a derivation).
Our example is easily generalized to the case of several independent
Wiener processes. See also \cite{DMH93-stoch}.
                                              \hfill  {\Large $\Box$}

\subsection{The case of a freely and finitely generated algebra}
Let $\cal A$ be freely generated by elements $x^i, \, i=1, \ldots,n,$
and the unit $\idty$.\footnote{$\cal A$
consists of finite linear combinations of monomials in $x^1, \ldots,
x^n$ and $\idty$ with coefficients in $\Cx$, i.e., ${\cal A} = \Cx
 \lbrack x^1, \ldots, x^n \rbrack $. Here, we will not discuss
a possible extension to infinite sums (e.g., the case of analytic
functions on $\Rl^n$).}
{}From Lemma 3.1 we can then deduce that the 1-forms
\be                          \label{taus}
   \tilde{\tau}^{i_1 \cdots i_r} := \tilde{\mbox{d}}x^{i_1} \bullet
   \cdots \bullet   \tilde{\mbox{d}}x^{i_r}     \qquad (r=1, \ldots)
\ee
generate $\tilde{\Omega}^1({\cal A})$ as a left $\cal A$-module.
$\tilde{\tau}^{i_1 \cdots i_r}$ is totally symmetric in the indices
$i_1, \ldots, i_r,$ so that we should restrict the latter by
$i_1 \leq i_2 \leq \ldots \leq i_r$.

\vskip.3cm
\noindent
{\bf Lemma 3.5} The set of 1-forms
\be
    B := \lbrace \tilde{\tau}^{i_1 \cdots i_r} \in
    \tilde{\Omega}^1({\cal A}) \mid i_1 \leq \dots \leq i_r, \,
     r=1,2,\dots \rbrace
\ee
is a basis of $\tilde{\Omega}^1({\cal A})$ as a left ${\cal
A}$-module.
\vskip.1cm
\noindent
{\bf Proof:} We already know that $B$ generates
$\tilde{\Omega}^1({\cal A})$ as a left ${\cal A}$-module. It is
therefore sufficient to show that any finite subset of $B$
is linearly independent over $\cal A$. Let
\begin{eqnarray*}
  0 = \sum_{r=1}^n \sum_{i_1 \leq i_2 \ldots \leq i_r}
      f_{i_1 \dots i_r} \, \tilde{\tau}^{i_1 \cdots i_r}
\end{eqnarray*}
with $f_{i_1 \dots i_r} \in {\cal A}$. Using
\begin{eqnarray*}
  \tilde{\tau}^{i_1 \cdots i_r} = \idty \otimes x^{i_1} \cdots
  x^{i_r} + \sum_{p=1}^{r-1} {(-1)^p \over p! (r-p)!} x^{(i_1}
  \cdots x^{i_p} \otimes x^{i_{p+1}} \cdots x^{i_r)}
  + (-1)^r x^{i_1} \cdots x^{i_r} \otimes \idty
\end{eqnarray*}
the above equation leads to
\begin{eqnarray*}
  0 = \sum_{i_1 \leq i_2 \ldots \leq i_n}
      f_{i_1 \dots i_n} \otimes x^{i_1} \cdots x^{i_n}
      + \mbox{rest}
\end{eqnarray*}
where `rest' consists of a finite sum of tensor products of which the
second factor is a monomial of degree $< n$ in the generators $x^i$ of
$\cal A$. Since $\cal A$ is freely generated, we conclude that
$f_{i_1 \dots i_n}=0$. By repetition of this argument,
$f_{i_1 \dots i_r} = 0 \; \forall i_1, \ldots, i_r, \, r=1, \ldots, n$.
\hspace*{\fill}                          {\Large $\Box$}

\vskip.3cm
\noindent
Similarly, one can argue that the
2-forms $\tilde{\tau}^{i_1 \cdots i_r} \, \tilde{\tau}^{i_1 \cdots i_s}$
($ r,s = 1, \ldots$) constitute a basis of $\tilde{\Omega}^2({\cal A})$
as a left $\cal A$-module (see also Lemma 3.2), and correspondingly for
$\tilde{\Omega}^r({\cal A})$ with $r>2$.
\vskip.3cm
\noindent
As a consequence of the preceding Lemma,
\be
 \tilde{\mbox{d}}f = \sum_{r=1}^\infty (\tilde{D}_{i_1 \cdots i_r} f)
       \, \tilde{\tau}^{i_1 \cdots i_r}  \; .  \label{df-D}
\ee
with operators $\tilde{D}_{i_1 \ldots i_r} \, : \, {\cal A}
\rightarrow {\cal A}$, where the indices are totally symmetric.
Inserted in (\ref{bull-Leibniz}) this leads to
\be
   \tilde{D}_{i_1 \cdots i_r}(f h) = f \, \tilde{D}_{i_1 \cdots i_r} h
      + h \, \tilde{D}_{i_1\cdots i_r} f +
      \sum_{k=1}^{r-1} {1 \over k! (r-k)!} \,
      (\tilde{D}_{(i_1 \cdots i_k} f) (\tilde{D}_{i_{k+1} \cdots i_r)} h)
\ee
which, in particular, shows that the operators $\tilde{D}_i$
are derivations. As a consequence of (\ref{df-D}) they satisfy
$\tilde{D}_j x^i = \delta^i_j$ and therefore coincide with
the ordinary partial derivatives\footnote{The `ordinary
partial derivatives' are the derivations $\partial_k \, : \,
{\cal A} \rightarrow {\cal A}$ ($k=1, \ldots, n$) with
$\partial_k x^\ell = \delta^\ell_k$.},
\be                            \label{D_i-part}
       \tilde{D}_i = \partial_i  \; .
\ee
Applying $\tilde{\mbox{d}}$ to (\ref{df-D}) using (\ref{dd-prod}) in
the form
\be
 \tilde{\mbox{d}} \tilde{\tau}^{i_1 \cdots i_r}
  & = & - \sum_{k=1}^{r-1}
 {1\over k! (r-k)!} \, \tilde{\tau}^{(i_1 \cdots i_k} \,
 \tilde{\tau}^{i_{k+1} \cdots i_r)}
\ee
and $\tilde{\mbox{d}}^2 = 0$, we obtain
\be
  \tilde{D}_{i_1 \cdots i_k} \, \tilde{D}_{i_{k+1} \cdots i_r} =
  {r \choose k} \, \tilde{D}_{i_1 \cdots i_r} \; .
\ee
Together with (\ref{D_i-part}) this implies
\be
 \tilde{D}_{i_1 \cdots i_r} = {1\over r!} \,
     \partial_{i_1} \cdots \partial_{i_r}  \qquad (r=1, \ldots) \; .
                                           \label{D-repr}
\ee
So far we have treated the case of the universal differential
calculus with its ring structure. For any other first order
differential calculus $(\Omega^1({\cal A}),\mbox{d})$ we can define
$\tau^{i_1 \cdots i_r}$ as in (\ref{taus}) (with $\tilde{\mbox{d}}x^i$
replaced by $\mbox{d}x^i$). These 1-forms are then, however, not
linearly independent. Nevertheless, the formulas derived above
induce corresponding formulas for any differential calculus, as
demonstrated in the following examples.
\vskip.3cm
\noindent
{\em Examples.} We evaluate (\ref{df-D}) with (\ref{D-repr}) for some
examples of differential calculi.
\begin{enumerate}
\item For the K\"ahler calculus, where $\mbox{d}x^i \bullet
    \mbox{d}x^j = 0$, we recover the familiar formula
    $\mbox{d}f = (\partial_i f) \, \mbox{d}x^i$.

\item The lattice calculus in \cite{DMHS93} is determined by
    $\mbox{d}x^i \bullet \mbox{d}x^j =  \ell \, \delta^{ij} \,
     \mbox{d}x^j$ (no summation, $\ell \in \Rl \setminus \lbrace 0 \rbrace$).
    In this case we obtain
\begin{eqnarray*}
    \mbox{d}f &=& \sum_{i=1}^n {1 \over \ell} \, ( \exp(\ell \partial_i)
                  -1) f \, \mbox{d}x^i      \\
    &=& \sum_{i=1}^n {1 \over \ell} \, \lbrack f(x^1,\ldots,x^i+\ell,
        \ldots,x^n) - f(x) \rbrack \, \mbox{d}x^i   \; .
\end{eqnarray*}
In the universal differential calculus, the ideal by which we have
to factorize $\tilde{\Omega}^1({\cal A})$ in order to obtain the
calculus under consideration is generated by
$\tilde{\mbox{d}}x^i \bullet \tilde{\mbox{d}}x^j -  \ell \, \delta^{ij}
 \, \tilde{\mbox{d}}x^j$.
Representing the $x^i$ as coordinate functions on $\Cx^n$ (or $\Rl^n$)
and evaluating the last expressions on $(a,b) \in \Cx^n \times \Cx^n$
using (\ref{univ-d}), we find
$$    (\tilde{\mbox{d}}x^i \bullet \tilde{\mbox{d}}x^j
      - \ell \, \delta^{ij} \, \tilde{\mbox{d}}x^j)(a,b)
     = (b^j - a^j) \lbrack b^i - a^i - \ell \delta^{ij} \rbrack \, .
$$
Equated to zero, this precisely displays the lattice structure.

\item For $n=1$ (for simplicity), the symmetric lattice calculus discussed
    in \cite{DMH94-discr} can be defined by
    $\, \mbox{d}x \bullet \mbox{d}x \bullet \mbox{d}x = \ell^2 \, \mbox{d}x
    \, $. Then
\begin{eqnarray*}
    \mbox{d}f &=& \sum_{r=0}^\infty {\ell^{2r} \over (2r+1)!} \,
    (\partial^{2r+1} f) \, \mbox{d}x + \sum_{r=1}^\infty {\ell^{2(r-1)}
    \over (2r)!} \, (\partial^{2r} f) \, \mbox{d}x \bullet \mbox{d}x  \\
    &=& \bar{\partial} f \, \mbox{d} x + {1 \over 2} \, \Delta f \,
        \mbox{d}x \bullet \mbox{d}x
\end{eqnarray*}
where
$$ \bar{\partial} f := { 1 \over 2 \ell} \, \lbrack f(x+\ell)
   - f(x-\ell) \rbrack \, , \quad
   \Delta f := {1 \over \ell^2} \, \lbrack f(x+\ell)+f(x-\ell)
   -2 \, f(x) \rbrack  \; .
$$
With $x$ as a coordinate function on $\Cx$, we find
$$   (\tilde{\mbox{d}}x \bullet \tilde{\mbox{d}}x \bullet
     \tilde{\mbox{d}}x - \ell^2 \, \tilde{\mbox{d}}x)(a,b)
     = (b - a) \lbrack (b - a)^2 - \ell^2 \rbrack  \; .
$$
Equated to zero, this implies $b=a$ or $b=a+\ell$ or $b=a-\ell$
which reveals the symmetric lattice structure (see \cite{DMH94-discr}).

For $\ell =0$ one obtains
$$ \mbox{d}f = \partial f \, \mbox{d}x + {1 \over 2} \,
   \partial^2 f \, \mbox{d}x \bullet \mbox{d}x  \; .
$$
The last type of calculus appears in the classical limit
($q \to 1$) of bicovariant differential calculus on the quantum
groups $SL_q(n)$ \cite{MH+Reut93,DMH93-stoch,Bres94}. Via
$ \mbox{d}x \bullet \mbox{d}x \mapsto \mbox{d}t $ contact is made
with the calculus of the last example in the previous subsection.

\item Generalizing the last two examples for $n=1$, we consider the
ideal in $\tilde{\Omega}^1({\cal A})$ generated by
$$
 (\tilde{\mbox{d}}x)^{\bullet (k+1)} - \ell^k \, \tilde{\mbox{d}}x
$$
for some fixed $k \in \Nl$. Evaluated on $(a,b) \in \Cx \times \Cx$,
it leads us to the equation
$$
     (b-a) \lbrack (b-a)^k - \ell^k \rbrack  = 0 \; .
$$
This defines an algorithm which, fixing a starting point $a$,
generates new points $a+ \ell q^r$ for $r=0,1, \ldots, k,$ where
$q$ is a primitive $k$th root of unity. In this way,
a lattice is created in the complex plane and the differential
calculus can be restricted to (the functions on) it.

Using $\sum_{j=0}^{k-1} q^j = (q^k -1)/(q-1) = 0$ for $k>1$, we find
\begin{eqnarray*}
 \mbox{d}f &=& \sum_{r=0}^\infty {\ell^{kr} \over (kr+1)!} \,
    (\partial^{kr+1} f) \, \mbox{d}x
  + \sum_{r=0}^\infty {\ell^{kr} \over (kr+2)!} \,
    (\partial^{kr+2} f) \, \mbox{d}x \bullet \mbox{d}x  \\
  & & + \ldots  + \sum_{r=0}^\infty {\ell^{kr}
    \over (kr+k)!} \, (\partial^{kr+k} f) \, (\mbox{d}x)^{\bullet k} \\
    &=& \sum_{j=1}^{k-1} (D_j f) \, (\mbox{d}x)^{\bullet j}
    + (D_k - \ell^{-k}) f \; (\mbox{d}x)^{\bullet k}
\end{eqnarray*}
with
$$
   D_j f =  {1 \over k \ell^j} \sum_{m=0}^{k-1} q^{j(k-m)} \,
            f(x+ \ell q^m)   \; .
$$
In terms of the 1-forms
$$
    \theta^j := {1 \over k} \sum_{m=1}^k q^{j(k-m)}
     \, \ell^{-m} \,(\mbox{d}x)^{\bullet m}
$$
this becomes
$$  \mbox{d}f = \sum_{j=0}^{k-1} \lbrack f(x+ \ell q^j) - f(x)
    \rbrack \, \theta^j   \; .
$$
Furthermore, we have the following simple commutation relations,
$$   \theta^j \, f(x) = f(x+\ell q^j) \, \theta^j  \; . $$

For $k=3$ the lattice is triangular and of the kind which underlies
the hard hexagon model in statistical mechanics \cite{Baxt82}.
More precisely, it should be regarded as an oriented lattice
and for $k=4$ one obtains the corresponding symmetric lattice.
The case $k=5$ is related to a quasilattice \cite{Jans88}.
\end{enumerate}
                                      \hfill  {\Large $\Box$}

\section{Deformations of the ordinary differential calculus on
         freely generated commutative algebras}
\setcounter{equation}{0}
Throughout this section $\cal A$ denotes an associative and commutative
algebra which is freely generated by $x^1, \ldots, x^n$ and the unit
$\idty$. Furthermore, we
restrict our considerations to $n$-{\em dimensional} first order
differential calculi $(\Omega^1({\cal A}), \mbox{d})$. For those the
differentials $\mbox{d}x^i, \, i=1, \ldots,n,$ form a basis of
$\Omega^1({\cal A})$ as a left $\cal A$-module. Such calculi may be
regarded as (algebraic) deformations of the ordinary (K\"ahler)
differential calculus and are therefore of particular interest.
(\ref{C-def}) then yields
\be          \label{dx-C}
   \lbrack \mbox{d}x^i , x^j \rbrack = C^{ij}{}{}_k \, \mbox{d}x^k
\ee
with $C^{ij}{}{}_k \in {\cal A}$. From (\ref{C-sym}) and the
Jacobi identity we obtain the following {\em consistency conditions},
\be                \label{C-conds}
 C^{ij}{}{}_k = C^{ji}{}{}_k \; , \quad
 C^{ik}{}{}_m \, C^{jm}{}{}_\ell = C^{jk}{}{}_m \, C^{im}{}{}_\ell
\ee
(see also \cite{DMHS93}). In terms of the (structure) matrices
${\bf C}^k$ with entries $({\bf C}^k)^i{}_j := C^{ki}{}{}_j$ the
first of these conditions means that the $j$th row of ${\bf C}^i$ equals
the $i$th row of ${\bf C}^j$. The second condition says that the
${\bf C}^i$ commute with each other:
\be                \label{C-comm}
     {\bf C}^k \, {\bf C}^\ell = {\bf C}^\ell \, {\bf C}^k   \; .
\ee
\vskip.3cm
\noindent
{\em Remark.} As described in the preceding section, a first order
differential calculus induces a product in the space of 1-forms. In
the case under consideration, the latter is determined by
\be                                \label{dxdx-prod}
  \mbox{d}x^i \bullet \mbox{d}x^j = C^{ij}{}{}_k \, \mbox{d}x^k \; .
\ee
As a consequence of (\ref{C-conds}) this product is commutative
and associative:
\be
       \mbox{d}x^i \bullet \mbox{d}x^j
     = C^{ij}{}_k \, \mbox{d}x^k
     = C^{ji}{}_k \, \mbox{d}x^k
     = \mbox{d}x^j \bullet \mbox{d}x^i
\ee
\be
     (\mbox{d}x^i \bullet \mbox{d}x^j) \bullet \mbox{d}x^k
 &=& (C^{ij}{}_\ell \, \mbox{d}x^\ell) \bullet \mbox{d}x^k
  = C^{ij}{}_\ell \, C^{\ell k}{}_m \, \mbox{d}x^m
  = C^{jk}{}_\ell \, C^{i\ell}{}_m \, \mbox{d}x^m  \nonumber \\
 &=& \mbox{d}x^i \bullet C^{jk}{}_\ell \, \mbox{d}x^\ell
  = \mbox{d}x^i \bullet (\mbox{d}x^j \bullet \mbox{d}x^k)    \; .
\ee
The matrices ${\bf C}^i$ constitute a representation of this algebra
since
\be
      {\bf C}^i {\bf C}^j = C^{ij}{}_k \, {\bf C}^k \; .
\ee
As a consequence of the foregoing, the classification of first order
differential calculi of the kind specified above with $C^{ij}{}{}_k
\in \Cx$ is equivalent to the classification of commutative and
associative algebras over $\Cx$.    \hfill {\Large $\Box$}

\vskip.3cm
\noindent
{\em Remark.} More generally, when the conditions (\ref{C-conds}) are
satisfied, (\ref{dx-C}) determines a (first order) differential calculus
on any algebra $\cal A$ which is freely generated by the $x^i$ modulo
commutation relations such that $\lbrack x^i , x^j \rbrack$ is constant
with respect to $\mbox{d}$ (for all $i,j$). Special examples are the
Heisenberg algebras of quantum mechanics (see also \cite{Herm77,DMH92}
for related work). Further examples are the algebras
considered in \cite{DFR94} where $\lbrack x^k , x^\ell \rbrack = i \,
Q^{k \ell}$ with an antisymmetric tensor operator $Q^{ij}$ which is
central in the algebra generated by the $x^k$.\footnote{Here we have to
make the assumption that $Q^{k \ell}$ is annihilated by $\mbox{d}$.}
The solutions of the consistency conditions presented
in subsections 4.3 and 4.4 therefore also determine differential calculi
on such noncommutative algebras.
                                             \hfill {\Large $\Box$}
\vskip.3cm
\noindent
In the following subsections we first introduce a notion of
`extension' of a differential calculus (following the general
receipe of section 2.4). A procedure for the classification
of differential calculi with constant structure functions is then
outlined and applied to the cases where $n=2$ and $n=3$.
The action of an exterior derivative on $\cal A$ is determined
by left- (or right-) partial derivatives, for which we derive
some general formulas and which we calculate for several
examples of differential calculi. Particular solutions of the
consistency conditions for arbitrary $n$ are discussed in the
last two subsections.

\subsection{Inner differential calculi and inner extensions of
            differential calculi}
The following result gives a criterion for a differential calculus
to be inner (in the sense of section 2.4).
\vskip.3cm
\noindent
{\bf Lemma 4.1} $\mbox{d}$ is inner if and only if there is a 1-form
$\rho = \rho_k \, \mbox{d}x^k$ with $\rho_k \, {\bf C}^k = {\bf 1}$
(the unit $n \times n$ matrix).
\vskip.1cm \noindent
{\bf Proof:} \\
``$\Rightarrow$'': follows immediately from (\ref{dx-C}) and
$\mbox{d}x^i = \lbrack \rho , x^i \rbrack$. \\
``$\Leftarrow$'':
\begin{eqnarray*}
 \mbox{d}f &=& \delta^i_j \, (D_i f) \, \mbox{d}x^j  \\
           &=& \rho_k \, (D_i f) \, C^{ik}{}_j \, \mbox{d}x^j
                                                            \\
           &=& \rho_k \, (D_i f) \, \lbrack \mbox{d}x^i , x^k
               \rbrack                                      \\
           &=& \rho_k \, \lbrack \mbox{d}f , x^k \rbrack    \\
           &=& \rho_k \, \lbrack \mbox{d}x^k , f \rbrack    \\
           &=& \lbrack \rho , f \rbrack    \; .
\end{eqnarray*}
                                               \hfill {\Large $\Box$}
\vskip.3cm
Let $(\Omega^1({\cal A}),\mbox{d})$ be a first order
differential calculus. To the generators
$x^1, \ldots, x^n$ of $\cal A$ we adjoin an element $x^{n+1}$ to
freely generate the larger commutative algebra
$\check{\cal A} = {\cal A} \lbrack x^{n+1} \rbrack$. On the
latter we introduce an $n$-dimensional first order differential
calculus via structure matrices as follows. Define
\be
  {\bf \check{C}}^i := \left( \begin{array}{c|c}
  {\bf C}^i  & \begin{array}{c} 0 \\  \vdots \\ 0 \\ \end{array} \\
  \hline e^i & 0 \\ \end{array} \right)
  \qquad (i=1, \ldots, n)
\ee
where ${\bf C}^i$ are the structure matrices of $(\Omega^1({\cal A}),
\mbox{d})$ and $e^i$ is the row vector with entries $e^i_j = \delta^i_j,
\, j=1, \ldots,n$. Let ${\bf \check{C}}^{n+1}$ be the $(n+1)\times (n+1)$
unit matrix. The matrices ${\bf \check{C}}^I, \, I=1, \ldots, n+1,$ then
satisfy the consistency conditions (\ref{C-conds}) (if the ${\bf C}^i$
satisfy them). For the enlarged differential calculus
$(\Omega^1(\check{\cal A}), \check{\mbox{d}}) =: \mbox{\bf Ext}
(\Omega^1({\cal A}),\mbox{d})$ the extended derivation is inner,
\be                            \label{df-inner}
  \check{\mbox{d}} f = \lbrack \check{\mbox{d}} x^{n+1} , f \rbrack
                     \qquad  (\forall f \in \check{\cal A}) \, ,
\ee
i.e., we have (\ref{df-inner-rho}) with $\rho = \check{\mbox{d}} x^{n+1}$.
In particular, if $(\Omega^1({\cal A}),\mbox{d})$ is not inner,
then there is always an extension of it which is inner. This
observation is helpful since it is often much easier to
carry out calculations with an inner exterior derivative.

\subsection{Procedure for classification of constant structure
            functions}
With the additional assumption that the structure functions are
constant, i.e., $C^{ij}{}{}_k \in \Cx$, it is in principle possible
to classify all first order differential calculi.\footnote{All first
order differential calculi with constant structure
functions extend to higher orders with the usual anticommutation
rule for the product of differentials,
$ \mbox{d} x^i \, \mbox{d} x^j = - \mbox{d} x^j \, \mbox{d} x^i $.
Of course, this simple rule does not extend to arbitrary 1-forms
in case of a noncommutative differential calculus (where some
of the $C^{ij}{}_k$ are different from zero).} This has been done
in \cite{DMHS93} for the case $n=2$. The methods used there are not
applicable to the case $n > 2$, however, in contrast to the procedure
which we outline below and which is then applied to the cases
$n=2$ and $n=3$.
\vskip.3cm

Under a $GL(n,\Cx)$-transformations
\be
          {x'}^k = U^k{}_\ell \, x^\ell
    \mbox{ with } U = (U^k{}_\ell) \in GL(n,\Cx)
\ee
the commutation relations (\ref{dx-C}) are invariant if
\be                   \label{C-transf}
 {C'}^{ij}{}_k = U^i{}_r \, U^j{}_s \, C^{rs}{}_t \, (U^{-1})^t{}_k
\ee
respectively,
\be                   \label{C'}
    {\bf C'}^i = U^i{}_j \, (U \, {\bf C}^j \, U^{-1})  \; .
\ee
These transformations preserve the conditions (\ref{C-conds}).
In order to classify differential calculi one should therefore
determine all equivalence classes of structure matrices with
respect to $GL(n,\Cx)$-transformations.
Thanks to the commutativity of the matrices ${\bf C}^i$, there is a
$U \in GL(n,\Cx)$ such that, for all $i =1, \ldots,n$, the
$U {\bf C}^i U^{-1}$ are triangular,
i.e., have zeros everywhere above the diagonal.
This is a consequence of the Jordan trigonalization theorem.
But then also the ${\bf C'}^i$ in (\ref{C'}) are triangular
as linear combinations of triangular matrices.
\vskip.3cm

Hence, in each $GL(n,\Cx)$-orbit of structure matrices there are
representatives which are triangular and only for those
we have to solve the $GL(n,\Cx)$-invariant conditions (\ref{C-conds}).
The symmetry condition now reduces the ${\bf C}^i$ to the following
form,
\be
   {\bf C}^1 = \left(
       \begin{array}{*{4}{c}}
       C^{11}{}_1  &  0       &  \cdots  &  0\\
       C^{12}{}_1  &  0       &  \cdots  &  0\\
       \vdots      &  \vdots  &          &  \vdots\\
       C^{1n}{}_1  &  0       &  \cdots  &  0\\
       \end{array}
               \right)   \, , \,
   {\bf C}^2 =  \left(
       \begin{array}{*{5}{c}}
       C^{21}{}_1  &  0           &  0      &  \cdots  &  0\\
       C^{22}{}_1  &  C^{22}{}_2  &  0      &  \cdots  &  0\\
       \vdots      &  \vdots      &  \vdots &          &  \vdots\\
       C^{2n}{}_1  &  C^{2n}{}_2  &  0      &  \cdots  &  0\\
       \end{array}
                \right)   \, ,  \dots   \nonumber \\
    \dots \, , \,
   {\bf C}^n   =  \left(
       \begin{array}{*{4}{c}}
       C^{n1}{}_1  &  0           &  \cdots  &  0\\
       C^{n2}{}_1  &  C^{n2}{}_2  &  \ddots  &  \vdots\\
       \vdots      &  \vdots      &  \ddots  &  0\\
       C^{nn}{}_1  &  C^{nn}{}_2  &  \cdots  &  C^{nn}{}_n\\
       \end{array} \right)    \; . \quad
       \label{C-structure}
\ee
Although we made use of the fact that the ${\bf C}^i$ have to commute
with each other in order to derive (\ref{C-structure}), the
commutativity is not yet built in completely.
The further procedure may now be as follows.
There are $GL(n,\Cx)$-transformations which preserve the above
form of the matrices ${\bf C}^i$. They can be used to further
simplify their structure. The remaining complex constants are then
constrained by quadratic equations resulting from the condition
(\ref{C-comm}) that the ${\bf C}^i$ have to commute with each other.
These equations have to be solved.
\vskip.3cm

In the simple case $n=1$, where $\lbrack \mbox{d}x , x \rbrack = c \,
\mbox{d}x$, there are two orbits. $c=0$ represents the ordinary
(K\"ahler) differential calculus. We refer to it
as \fbox{\bf K}. The other orbit where $c \neq 0$ can be represented
by $c=1$. It describes a differential calculus on a 1-dimensional
lattice\footnote{To see this, one actually has to go beyond the
algebra of polynomials since functions with period $c$ play an
essential role in this case \cite{DMHS93}.} denoted by \fbox{\bf L}.
\vskip.3cm

{}From these 1-dimensional calculi one can build differential
calculi on algebras with more than one generator. The general
construction has been recalled in section 2.3.
Let $y^1,\ldots,y^r$ and $z^1,\ldots,z^s$ be the generators
of two commutative algebras with, respectively, $r$- and $s$-dimensional
(first order) differential calculi determined by
\be
 \lbrack \mbox{d} y^a , y^b \rbrack = C^{ab}{}{}_c \, \mbox{d}y^c
       \, , \quad
 \lbrack \mbox{d} z^{a'} , z^{b'} \rbrack = C^{a'b'}{}{}_{c'} \,
      \mbox{d}z^{c'}  \; .
\ee
For $x^a := y^a \otimes \idty$ and $x^{r+a'} := \idty \otimes z^{a'}$
this implies
\be
 \lbrack \mbox{d} x^i , x^j \rbrack = \hat{C}^{ij}{}{}_k \, \mbox{d}x^k
          \qquad (i,j = 1, \ldots, r+s)
\ee
where $\hat{C}^{ab}{}{}_c = C^{ab}{}{}_c$, $\hat{C}^{r+a', r+b'}{}{}
_{r+c'} = C^{a'b'}{}{}_{c'}$ and $\hat{C}^{ij}{}{}_k =0$ otherwise.
Conversely, if after some $GL(n,\Cx)$-transformation the structure
matrices of a differential calculus decompose in this way, the calculus
is reducible and can be expressed as a skew tensor product of
lower-dimensional calculi.
\vskip.3cm

The $n$-dimensional irreducible calculi can be further classified into
those which are extensions -- in the sense of subsection 4.1 -- of
$(n-1)$-dimensional calculi and those which are not. This makes sense
on the basis of the following result.
\vskip.3cm
\noindent
{\bf Lemma 4.2} The extensions of all representatives of a
$GL(n,\Cx)$-orbit of $n$-dimensional differential calculi
lie in the same $GL(n+1,\Cx)$-orbit.
\vskip.1cm
\noindent
{\bf Proof:} With the special $GL(n+1,\Cx)$-matrix
\begin{eqnarray*}
   \check{U}=\left(\begin{array}{cc} U & 0 \\
                                   0 & 1 \end{array}\right)
\end{eqnarray*}
where $U \in GL(n,\Cx)$ we find
\begin{eqnarray*}
  \check{\bf C}'{}^i &=& \check{U}^i{}_J \, \check{U} \check{\bf C}^J
                       \check{U}^{-1}
                    = U^i{}_j \left(\begin{array}{cc}
                          U & 0 \\
                          0 & 1 \end{array} \right)
  \left(\begin{array}{cc} {\bf C}^j & 0 \\
                          e^j & 0 \end{array}\right)
  \left(\begin{array}{cc} U^{-1} & 0 \\
                           0      & 1 \end{array}\right)   \\
  &=& \left(\begin{array}{cc}
                          U^i{}_j \, U {\bf C}^j U^{-1}  & 0 \\
                          U^i{}_j \, e^j \, U^{-1} & 0
                          \end{array} \right)  \; .
\end{eqnarray*}
Since $U^i{}_j \, e^j \, U^{-1} = e^i$, this is the extension of the
$U$-transformed ${\bf C}^i$. Furthermore, $\check{\bf C}'{}^{n+1}
= \check{\bf C}^{n+1}$ (which is the $(n+1) \times (n+1)$ unit matrix).
                                              \hfill {\Large $\Box$}

\subsection{Classification of 2-dimensional differential calculi}
For $n=2$ equation (\ref{C-structure}) becomes
\be
   {\bf C}^1   =  \left(
      \begin{array}{*{2}{c}}
      a  &  0\\
      b  &  0
       \end{array} \right)     \, , \,
   {\bf C}^2   =  \left(
      \begin{array}{*{2}{c}}
      b  &  0\\
      c  &  d
       \end{array} \right)  \; .      \label{C-2d}
\ee
The two matrices commute iff the complex constants $a,b,c,d$ are related
by
\be                     \label{abcd-rel}
          b^2 - a \, c - b \, d = 0  \, .
\ee
An arbitrary element of $GL(2,\Cx)$ is given by
\be
      \left ( \begin{array}{*{2}{c}}
              s  &  t \\
              u  &  v
              \end{array} \right)
\ee
with ${\cal D} := sv-tu \neq 0$. It acts on the matrices ${\bf C}^i$ as
follows,
\be
\begin{array}{l}
 {\bf C'}^1 = {1 \over {\cal D}} \, \left ( \begin{array}{*{2}{c}}
  s^2 v a + 2 s t v b + t^2 v c -t^2 u d & -s^2 t a - 2 s t^2 b
                      -t^3 c + s t^2 d  \\
  s u v a + v (s v + t u) b + t v^2 c - t u v d  &
  -s t u a - t (s v + t u) b - t^2 v c + s t v d
              \end{array} \right)       \\
 {\bf C'}^2 = {1 \over {\cal D}} \, \left ( \begin{array}{*{2}{c}}
  s u v a + v (t u + s v) b +t v^2 c -t u v d & -s t u a
       - t (t u + s v) b -t^2 v c + s t v d  \\
  u^2 v a + 2 u v^2 b + v^3 c - u v^2 d  &
  -t u^2 a - 2 t u v b - t v^2 c + s v^2 d
              \end{array} \right)  \,
\end{array}
              \label{C-transf2}
\ee
For $t=0$ this transformation preserves the form of the matrices in
(\ref{C-2d}),
\be
 {\bf C'}^1 = \left ( \begin{array}{*{2}{c}}
              s a & 0  \\
              u a + v b  & 0
              \end{array} \right)  \, , \,
 {\bf C'}^2 = \left ( \begin{array}{*{2}{c}}
              u a + v b & 0  \\
              {1 \over s} (u^2 a + 2 uvb + v^2 c - uvd) & v d
              \end{array} \right)  \; .
\ee
It can thus be used to further reduce the parameter freedom
of the matrices ${\bf C}^i$.
\vskip.3cm
\noindent
If $a \neq 0$, we set $s = 1/a$ and $u = - vb/a$. Then
\be
 {\bf C'}^1 = \left ( \begin{array}{*{2}{c}}
              1 & 0  \\
              0  & 0
              \end{array} \right)  \, , \,
 {\bf C'}^2 = \left ( \begin{array}{*{2}{c}}
              0 & 0  \\
              0 & v d
              \end{array} \right)
\ee
using (\ref{abcd-rel}). If $d=0$ we have ${\bf C'}^2 =0$. Otherwise
the choice $v = 1/d$ leads to
\be           \label{C2_dr1}
 {\bf C'}^2 = \left ( \begin{array}{*{2}{c}}
              0 & 0  \\
              0 & 1
              \end{array} \right)  \; .
\ee
If $a = 0$ and $b=0$, so that ${\bf C'}^1 =0$, we can arrange
either ${\bf C'}^2 =0$, ${\bf C'}^2$ of the form (\ref{C2_dr1}), or
\be
 {\bf C'}^2 = \left ( \begin{array}{*{2}{c}}
              0 & 0  \\
              1 & 0
              \end{array} \right)  \; .
\ee
In the remaining case $a=0$ and $b \neq 0$ one can always reach
\be
 {\bf C'}^1 = \left ( \begin{array}{*{2}{c}}
              0 & 0  \\
              1 & 0
              \end{array} \right)  \, , \,
 {\bf C'}^2 = \left ( \begin{array}{*{2}{c}}
              1 & 0  \\
              0 & 1
              \end{array} \right)  \; .
\ee
In all these cases, (\ref{abcd-rel}) is automatically satisfied.
It has still to be checked, with the help of (\ref{C-transf2}),
which of the representatives for ${\bf C}^1$ and ${\bf C}^2$ obtained
in this way generate different orbits. In the following we list
representatives from all distinct orbits. The respective complete orbit
is then obtained via (\ref{C-transf2}).\footnote{The pair of matrices
${\bf C}^i$ with $a=1$ and $b=c=d=0$ which we encountered above lies in
the orbit of solution {\bf (2)}.}
\vskip.3cm
\noindent

\renewcommand{\labelenumi}{\bf (\arabic{enumi})}
\begin{enumerate}

\item For ${\bf C}^1 = {\bf C}^2 = 0$ we recover the commutative
(K\"ahler) differential calculus. It is reducible since it is the
skew tensor product of two 1-dimensional commutative differential
calculi:  \fbox{\bf K}$\hat{\otimes}$\fbox{\bf K}.

\item The pair of matrices
\begin{eqnarray*}
   {\bf C}^1   =  \left(\begin{array}{cc}
                        0  &  0 \\
                        0  &  0
                  \end{array} \right)  \, , \,
   {\bf C}^2   =  \left(\begin{array}{cc}
                        0  &  0 \\
                        0  &  1
                  \end{array} \right)
\end{eqnarray*}
represents  \fbox{\bf K}$\hat{\otimes}$\fbox{\bf L}.

\item The matrix pair
\begin{eqnarray*}
   {\bf C}^1   =  \left(\begin{array}{cc}
                        1  &  0 \\
                        0  &  0
                  \end{array} \right)  \, , \,
   {\bf C}^2   =  \left(\begin{array}{cc}
                        0  &  0 \\
                        0  &  1
                  \end{array} \right)
\end{eqnarray*}
corresponds to \fbox{\bf L}$\hat{\otimes}$\fbox{\bf L}
 = {\bf Ext}(\fbox{\bf L}).

\item A further calculus is given by
\begin{eqnarray*}
   {\bf C}^1   =  \left(\begin{array}{cc}
                        0  &  0 \\
                        0  &  0
                        \end{array} \right) \, , \,
   {\bf C}^2   =  \left(\begin{array}{cc}
                        0  &  0  \\
                        1  &  0
                        \end{array} \right) \; .
\end{eqnarray*}
It is neither reducible nor the extension of a 1-dimensional calculus.
It therefore plays a role as a `building block' for the construction of
higher-dimensional differential calculi. We will refer to it via
\framebox[1cm]{\bf I}. This calculus is a special case of a class
of calculi which has been investigated in \cite{DMH92-grav,DMH93-stoch}.

\item Another irreducible calculus is determined by
\begin{eqnarray*}
     {\bf C}^1   =  \left(\begin{array}{cc}
                          0  &  0 \\
                          1  &  0
                    \end{array} \right) \, , \,
     {\bf C}^2   =  \left(\begin{array}{cc}
                          1  &  0 \\
                          0  &  1
                    \end{array} \right)  \; .
\end{eqnarray*}
It is the extension of \fbox{\bf K} and shall hence be denoted as
{\bf Ext}(\fbox{\bf K}).
\end{enumerate}

\vskip.3cm
\noindent
If we want to have an involution on the differential algebra,
we have to decompose the calculi into orbits with respect to the
action of $GL(2,\Rl)$. The $GL(2,\Cx)$-orbit of {\bf (3)} then splits
into two $GL(2,\Rl)$-orbits (cf \cite{DMHS93}).

\subsection{Classification of 3-dimensional differential calculi}
In this subsection we apply the procedure described in subsection 4.2
to the case of an algebra with three generators. (\ref{C-structure})
then reads
\be
   {\bf C}^1   =  \left(\begin{array}{ccc}
                         a  &  0  &  0 \\
                         b  &  0  &  0 \\
                         c  &  0  &  0
                  \end{array} \right)   \, , \,
   {\bf C}^2   =  \left(\begin{array}{ccc}
                         b  &  0  &  0 \\
                         d  &  e  &  0 \\
                         f  &  g  &  0
                  \end{array} \right)     \, , \,
   {\bf C}^3   =  \left(\begin{array}{*{3}{c}}
                         c  &  0  &  0 \\
                         f  &  g  &  0 \\
                         h  &  k  &  l
                  \end{array} \right)   \; .
\ee
The complex entries are subject to the relations
\be
\begin{array}{lcc}
 b^2 - be - ad             & = & 0 \\
 bc - af - bg              & = & 0 \\
 c^2 - ah - bk - cl        & = & 0 \\
 cd + ef - bf - dg         & = & 0 \\
 cf + fg - bh - dk - fl    & = & 0 \\
 g^2 - ek - gl             & = & 0
\end{array}
\ee
Proceeding as in the 2-dimensional case treated in the previous
subsection, after a tedious calculation one ends up with the following
list of representatives of $GL(3,\Cx)$-orbits.\footnote{Some more
details are presented in \cite{Baeh94}.}
\begin{enumerate}
\item
\fbox{\bf K}$\hat{\otimes}$\fbox{\bf K}$\hat{\otimes}
$\fbox{\bf K}

\item
\fbox{\bf K}$\hat{\otimes}$\fbox{\bf K}$\hat{\otimes}
$\fbox{\bf L}
\begin{eqnarray*}
   {\bf C}^1 = \left(\begin{array}{*{3}{c}}
                     0  &  0  &  0 \\
                     0  &  0  &  0 \\
                     0  &  0  &  0
               \end{array} \right)  \, , \,
   {\bf C}^2 = \left(\begin{array}{*{3}{c}}
                     0  &  0  &  0 \\
                     0  &  0  &  0 \\
                     0  &  0  &  0
               \end{array} \right) \, , \,
   {\bf C}^3 = \left(\begin{array}{*{3}{c}}
                     0  &  0  &  0 \\
                     0  &  0  &  0 \\
                     0  &  0  &  1
               \end{array} \right) \, .
\end{eqnarray*}

\item
\fbox{\bf K}$\hat{\otimes}$\fbox{\bf L}$\hat{\otimes}
$\fbox{\bf L}
\begin{eqnarray*}
   {\bf C}^1 = \left(\begin{array}{*{3}{c}}
                     0  &  0  &  0 \\
                     0  &  0  &  0 \\
                     0  &  0  &  0 \\
               \end{array} \right)  \, ,  \,
   {\bf C}^2 = \left(\begin{array}{*{3}{c}}
                     0  &  0  &  0 \\
                     0  &  1  &  0 \\
                     0  &  0  &  0
               \end{array} \right) \, , \,
   {\bf C}^3 = \left(\begin{array}{*{3}{c}}
                     0  &  0  &  0 \\
                     0  &  0  &  0 \\
                     0  &  0  &  1
               \end{array} \right)  \; .
\end{eqnarray*}

\item
\fbox{\bf L}$\hat{\otimes}$\fbox{\bf L}$\hat{\otimes}
$\fbox{\bf L}
 = {\bf Ext}(\fbox{\bf L}$\hat{\otimes}$\fbox{\bf L})
\begin{eqnarray*}
   {\bf C}^1 = \left(\begin{array}{*{3}{c}}
                     1  &  0  &  0 \\
                     0  &  0  &  0 \\
                     0  &  0  &  0 \\
               \end{array} \right) \, , \,
   {\bf C}^2 = \left(\begin{array}{*{3}{c}}
                     0  &  0  &  0 \\
                     0  &  1  &  0 \\
                     0  &  0  &  0
               \end{array} \right)  \, , \,
   {\bf C}^3 = \left(\begin{array}{*{3}{c}}
                     0  &  0  &  0 \\
                     0  &  0  &  0 \\
                     0  &  0  &  1
               \end{array} \right)  \; .
\end{eqnarray*}

\item
\framebox[1cm]{\bf I}$\hat{\otimes}$\fbox{\bf K}
\begin{eqnarray*}
   {\bf C}^1 = \left(\begin{array}{*{3}{c}}
                     0  &  0  &  0 \\
                     0  &  0  &  0 \\
                     0  &  0  &  0
               \end{array} \right)  \, , \,
   {\bf C}^2 = \left(\begin{array}{*{3}{c}}
                     0  &  0  &  0 \\
                     1  &  0  &  0 \\
                     0  &  0  &  0
               \end{array} \right)  \, , \,
   {\bf C}^3 = \left(\begin{array}{*{3}{c}}
                     0  &  0  &  0 \\
                     0  &  0  &  0 \\
                     0  &  0  &  0
               \end{array} \right) \; .
\end{eqnarray*}

\item
\framebox[1cm]{\bf I}$\hat{\otimes}$\fbox{\bf L}
\begin{eqnarray*}
   {\bf C}^1 = \left(\begin{array}{*{3}{c}}
                     0  &  0  &  0 \\
                     0  &  0  &  0 \\
                     0  &  0  &  0
               \end{array} \right)  \, , \,
   {\bf C}^2 = \left(\begin{array}{*{3}{c}}
                     0  &  0  &  0 \\
                     1  &  0  &  0 \\
                     0  &  0  &  0
               \end{array} \right)  \, , \,
   {\bf C}^3 = \left(\begin{array}{*{3}{c}}
                     0  &  0  &  0 \\
                     0  &  0  &  0 \\
                     0  &  0  &  1
               \end{array} \right)  \; .
\end{eqnarray*}

\item
{\bf Ext}(\fbox{\bf K})$\hat{\otimes}$\fbox{\bf K}
\begin{eqnarray*}
   {\bf C}^1 = \left(\begin{array}{*{3}{c}}
                     0  &  0  &  0  \\
                     1  &  0  &  0  \\
                     0  &  0  &  0
               \end{array} \right)  \, , \,
   {\bf C}^2 = \left(\begin{array}{*{3}{c}}
                     1  &  0  &  0  \\
                     0  &  1  &  0  \\
                     0  &  0  &  0
               \end{array} \right)  \, , \,
   {\bf C}^3 = \left(\begin{array}{*{3}{c}}
                     0  &  0  &  0  \\
                     0  &  0  &  0  \\
                     0  &  0  &  0
               \end{array} \right)  \; .
\end{eqnarray*}

\item
{\bf Ext}(\fbox{\bf K})$\hat{\otimes}$\fbox{\bf L}
 = {\bf Ext}({\bf Ext}(\fbox{\bf K}))
 = {\bf Ext}(\fbox{\bf K}$\hat{\otimes}$\fbox{\bf L})
\begin{eqnarray*}
   {\bf C}^1 = \left(\begin{array}{ccc}
                     0  &  0  &  0 \\
                     1  &  0  &  0 \\
                     0  &  0  &  0
               \end{array} \right)  \, ,  \,
   {\bf C}^2 = \left(\begin{array}{ccc}
                     1  &  0  &  0 \\
                     0  &  1  &  0 \\
                     0  &  0  &  0
               \end{array} \right)  \, , \,
   {\bf C}^3 = \left(\begin{array}{ccc}
                     0  &  0  &  0 \\
                     0  &  0  &  0 \\
                     0  &  0  &  1
               \end{array} \right)  \; .
\end{eqnarray*}

\item
An irreducible calculus is given by
\begin{eqnarray*}
   {\bf C}^1 = \left(\begin{array}{*{3}{c}}
                     0  &  0  &  0  \\
                     0  &  0  &  0  \\
                     0  &  0  &  0
               \end{array} \right)  \, , \,
   {\bf C}^2 = \left(\begin{array}{*{3}{c}}
                     0  &  0  &  0 \\
                     0  &  0  &  0 \\
                     1  &  0  &  0
               \end{array} \right)  \, , \,
   {\bf C}^3 = \left(\begin{array}{*{3}{c}}
                     0  &  0  &  0 \\
                     1  &  0  &  0 \\
                     0  &  0  &  0
               \end{array} \right) \; .
\end{eqnarray*}

\item
Another irreducible calculus is determined by
\begin{eqnarray*}
   {\bf C}^1 = \left(\begin{array}{*{3}{c}}
                     0  &  0  &  0 \\
                     0  &  0  &  0 \\
                     0  &  0  &  0
               \end{array} \right)  \, , \,
   {\bf C}^2 = \left(\begin{array}{*{3}{c}}
                     0  &  0  &  0 \\
                     0  &  0  &  0 \\
                     1  &  0  &  0
               \end{array} \right)  \, , \,
   {\bf C}^3 = \left(\begin{array}{*{3}{c}}
                     0  &  0  &  0 \\
                     1  &  0  &  0 \\
                     0  &  1  &  0
               \end{array} \right)  \; .
\end{eqnarray*}

\item
{\bf Ext}(\fbox{\bf K}$\hat{\otimes}$\fbox{\bf K})
\begin{eqnarray*}
   {\bf C}^1 = \left(\begin{array}{*{3}{c}}
                     0  &  0  &  0 \\
                     0  &  0  &  0 \\
                     1  &  0  &  0
               \end{array} \right)  \, , \,
   {\bf C}^2 = \left(\begin{array}{*{3}{c}}
                     0  &  0  &  0 \\
                     0  &  0  &  0 \\
                     0  &  1  &  0
               \end{array} \right)  \, , \,
   {\bf C}^3 = \left(\begin{array}{*{3}{c}}
                     1  &  0  &  0 \\
                     0  &  1  &  0 \\
                     0  &  0  &  1
               \end{array} \right)  \; .
\end{eqnarray*}

\item
{\bf Ext}(\framebox[1cm]{\bf I})
\begin{eqnarray*}
   {\bf C}^1 = \left(\begin{array}{*{3}{c}}
                     0  &  0  &  0 \\
                     0  &  0  &  0 \\
                     1  &  0  &  0
               \end{array} \right) \, , \,
   {\bf C}^2 = \left(\begin{array}{*{3}{c}}
                     0  &  0  &  0 \\
                     1  &  0  &  0 \\
                     0  &  1  &  0
               \end{array} \right)  \, , \,
   {\bf C}^3 = \left(\begin{array}{*{3}{c}}
                     1  &  0  &  0 \\
                     0  &  1  &  0 \\
                     0  &  0  &  1
               \end{array} \right)  \; .
\end{eqnarray*}

\end{enumerate}

The last four of these calculi are irreducible.
Only two calculi -- {\bf (9)} and {\bf (10)} -- are new in the sense
that they cannot be obtained as a skew tensor product or an extension
of lower-dimensional calculi. We shall see in subsection 4.5 that
{\bf (9)} is a special case of the calculus explored in
\cite{DMH92-grav,DMH93-stoch} for arbitrary $n$, to which also the
2-dimensional calculus \framebox[1cm]{\bf I} belongs. A generalization
of {\bf (10)} to arbitrary $n$ is presented in subsection 4.6.

\subsection{Left- and right-partial derivatives}
{\em Left-partial derivatives} are defined as $\Cx$-linear maps
$D_j \, : \, {\cal A} \rightarrow {\cal A}$ by
\be
 \mbox{d} f =: (D_j f) \, \mbox{d} x^j
       \qquad \forall f \in {\cal A}    \; .
\ee
Using (\ref{dx-C}), one finds
\be
    \lbrack \mbox{d} f , h \rbrack
 &=& (D_i f) \, \lbrack \mbox{d} x^i , h \rbrack
  = (D_i f) \, \lbrack \mbox{d} h , x^i \rbrack
  = (D_i f) (D_j h) \, \lbrack \mbox{d} x^j , x^i \rbrack
       \nonumber \\
 &=& (D_i f) (D_j h) \, C^{ij}{}_k \, \mbox{d} x^k  \; .
\ee
This leads to
\be
 D_j (fh) \, dx^j &=& \mbox{d} (fh) = f \, \mbox{d} h
     + h \, \mbox{d}f + \lbrack \mbox{d} f , h \rbrack  \nonumber \\
 &=& \lbrace (D_j f) \, h + f \, D_j h +
     (D_k f) (D_\ell h) \, C^{k \ell}{}_j \rbrace \,
     \mbox{d} x^j              \qquad \forall f,h \in {\cal A}
     \label{twist-Leibniz}
\ee
from which we can read off a twisted Leibniz rule for the $D_j$.
\vskip.3cm
\noindent
{\bf Lemma 4.3} The left-partial derivatives are given by
\be                          \label{part-sum}
   D_j =  \sum_{r=1}^\infty \frac{1}{r!} \,
              ({\bf C}^{k_1} \cdots {\bf C}^{k_{r-1}})^{k_r}{}_j \;
              \partial_{k_1} \dots \partial_{k_r}
\ee
in terms of ordinary partial derivatives.\footnote{The first summand
on the rhs is $\partial_j$.}
\vskip.1cm \noindent
{\bf Proof:}
First we note that $({\bf C}^{k_1} \cdots {\bf C}^{k_{r-1}})^{k_r}{}_j$
is totally symmetric in the indices $k_1, \ldots, k_r$ as a
consequence of $C^{ij}{}_k = C^{ji}{}_k$ and the commutativity of
the structure matrices ${\bf C}^i$.
Because of the $\Cx$-linearity of the $D_j$ it is sufficient
to prove (\ref{part-sum}) on monomials in $x^i, \, i=1, \ldots,n$.
This will be done using induction with respect to the degree of
monomials. Applied to $x^i$ the formula is obviously true. Let us
assume that it holds for monomials up to degree $m$. If $u$ is
a monomial of degree $m$, then
\begin{eqnarray*}
 & &  \sum_{r=1}^\infty {1 \over r!} \, ({\bf C}^{k_1} \cdots
     {\bf C}^{k_{r-1}})^{k_r}{}_j \, \partial_{k_1} \cdots
     \partial_{k_r} (x^i u)   \\
 &=& \partial_j (x^i u) + \sum_{r=2}^\infty {1 \over (r-1)!} \,
     ({\bf C}^{k_1} \cdots {\bf C}^{k_{r-1}})^i{}_j \, \partial_{k_1}
     \cdots \partial_{k_{r-1}} u  \\
 & & + x^i \,  \sum_{r=2}^\infty {1 \over r!} \, ({\bf C}^{k_1} \cdots
     {\bf C}^{k_{r-1}})^{k_r}{}_j \, \partial_{k_1} \cdots
     \partial_{k_r} u   \\
 &=& \delta^i_j \, u + \sum_{r=2}^\infty {1 \over (r-1)!} \,
     ({\bf C}^{k_1} \cdots {\bf C}^{k_{r-2}})^{k_{r-1}}{}_\ell \,
     C^{i \ell}{}_j \, \partial_{k_1} \cdots \partial_{k_{r-1}} u
     + x^i \, D_j u                        \\
 &=& \delta^i_j \, u + C^{i \ell}{}_j \, \sum_{r=1}^\infty {1 \over r!}
     \, ({\bf C}^{k_1} \cdots {\bf C}^{k_{r-1}})^{k_r}{}_\ell \,
     \partial_{k_1} \cdots \partial_{k_r} u + x^i \, D_j u \\
 &=& \delta^i_j \, u + C^{i \ell}{}_j \, D_\ell u
     + x^i \, D_j u           \\
 &=& (D_j x^i) \, u + x^i \, D_j \, u
     + C^{k \ell}{}_j \, (D_k x^i) \, D_\ell u
  = D_j (x^i u)      \; .
\end{eqnarray*}
where we used (\ref{twist-Leibniz}) and $D_j x^i = \delta^i_j$
in the last steps.                             \hfill  {\Large $\Box$}
\vskip.3cm
\noindent
{\em Remark.} For any first order differential calculus
$(\Omega^1({\cal A}),\mbox{d})$ we define $\tau^{i_1 \cdots i_r}$ as
in (\ref{taus}) (with $\tilde{\mbox{d}}x^i$ replaced by $\mbox{d}x^i$).
Then (\ref{dxdx-prod}) leads to
\be
   \tau^{i_1 \cdots i_r} = C^{i_1 i_2}{}_{k_1}
   C^{i_3 k_1}{}_{k_2} \cdots C^{i_r k_{r-2}}{}_\ell \,
   \mbox{d}x^\ell  \; .
\ee
Inserting this in (\ref{df-D}), we obtain
\be
   \mbox{d}f = D_k f \, \mbox{d} x^k
\ee
with
\be
   D_k = \partial_k +\sum_{r=2}^\infty
  {1\over r!} \, C^{i_1 i_2}{}_{j_1} C^{i_3 j_1}{}_{j_2}
  \cdots C^{i_r j_{r-2}}{}_k \; \partial_{i_1} \cdots
  \partial_{i_r}  \; .
\ee
This is our formula (\ref{part-sum}).     \hfill {\Large $\Box$}

\vskip.3cm
\noindent
{\bf Lemma 4.4}
\be                             \label{dx^i-f-comm}
 \mbox{d} x^i \, f = \left. \exp \left( {\bf C}^k(x) \, {\partial \over
  \partial y^k} \right)^i_j \, f(y) \right |_{y=x}  \; \mbox{d} x^j \;.
\ee
\vskip.1cm \noindent
{\bf Proof:} We have
\begin{eqnarray*}
  \lbrack \mbox{d}x^i , f \rbrack = \lbrack \mbox{d} f , x^i \rbrack
  = (D_j f) \, \lbrack \mbox{d} x^j , x^i \rbrack
  = (D_j f) \, C^{ij}{}_k \, \mbox{d} x^k  \; .
\end{eqnarray*}
Inserting the expression (\ref{part-sum}) for $D_j$, we find
\begin{eqnarray*}
     \lbrack \mbox{d}x^i , f \rbrack
 &=& \sum_{r=1}^\infty \frac{1}{r!} \, ({\bf C}^{k_1} \dots
    {\bf C}^{k_{r-1}} {\bf C}^i)^{k_r}{}_j \,
    \partial_{k_1} \dots \partial_{k_r} f \; \mbox{d} x^j \\
 &=& \sum_{r=1}^\infty \frac{1}{r!} \, ({\bf C}^{k_1} \dots
    {\bf C}^{k_r})^i{}_j \, \partial_{k_1} \dots \partial_{k_r} f \;
    \mbox{d} x^j \\
 &=& \left. \left( \exp \left( {\bf C}^k(x) \, {\partial \over
  \partial y^k} \right) - {\bf 1} \right)^i_j f(y) \right |_{y=x}
  \; \mbox{d} x^j   \;.
\end{eqnarray*}
Here we have stressed the possible $x^k$-dependence of the structure
matrices which necessitates the introduction of the auxiliary variables
$y^k$ in the last formula.                    \hfill {\Large $\Box$}

\vskip.3cm

Let us suppose that $\Omega^1({\cal A})$ considered as an algebra
with the product $\bullet$ (see section 3) is nilpotent, i.e., there
is a number $k \in \Nl \setminus \{0\}$ such that all products with $k$
faktors vanish (see \cite{Abia71}, for example).
The smallest such number is called the {\em index} of the algebra.
Since multiplication is determined through the $n\times n$
matrices ${\bf C}^i$, the index can be maximally $n$. Then
(\ref{part-sum}) shows that the left-partial derivatives
are differential operators of at most $n$th order.

If $(\Omega^1({\cal A}), \bullet)$ is not nilpotent, then there is a
nonvanishing idempotent element (see \cite{Abia71}, for example).
The sum in (\ref{part-sum}) is then not finite, so that the
left-partial derivatives are non-local. This is the case, for
example, for the `lattice calculus' \fbox{\bf L} and for each
differential calculus which is an extension in the sense of
section 4.1.
\vskip.3cm

For the {\em right-partial derivatives} $D_{-j}$ defined
by $\mbox{d}f =: \mbox{d}x^j \, D_{-j} f$ the formula
(\ref{part-sum}) is replaced by
\be
   D_{-j} =  \sum_{r=1}^\infty \frac{(-1)^{r-1}}{r!} \,
              ({\bf C}^{k_1} \cdots {\bf C}^{k_{r-1}})^{k_r}{}_j \;
              \partial_{k_1} \dots \partial_{k_r}  \; .
\ee
\vskip.3cm
\noindent
{\em Examples.}
In the following we examine the four irreducible calculi which we
found in subsection 4.4 and present the corresponding left- and
right-partial derivatives.
\begin{enumerate}
\setcounter{enumi}{8}
\item In this case, the only nonvanishing commutators (\ref{dx-C})
      are
  \be
     \lbrack \mbox{d}x^2 , x^3 \rbrack = \lbrack \mbox{d}x^3 , x^2
     \rbrack = \mbox{d}x^1  \; .
  \ee
  The corresponding left- and right-partial derivatives are
  \be
  D_{\pm 1} = \partial_1 \pm \partial_2 \partial_3 , \quad
  D_{\pm 2} = \partial_2 , \quad
  D_{\pm 3} = \partial_3
  \ee
  in accordance with the fact that the index of the associated
  algebra is equal to 2.
  In terms of $y^1 := x^1$, $y^2 := {i \over \sqrt{2}} \, (x^2 + x^3)$
  and $y^3 := {1 \over \sqrt{2}} \, (x^2 - x^3)$ we obtain
  \be
     \lbrack \mbox{d}y^\mu , y^\nu \rbrack = - \delta^{\mu \nu} \,
     \mbox{d}y^1            \qquad  (\mu,\nu = 2,3)    \; .
  \ee
  This is a special case of a differential calculus which has been
  studied in \cite{DMH92-grav,DMH93-stoch} (see also (\ref{ex-grav})
  and the second example in section 3.2).
\item Here, the nonvanishing commutators are
  \be
     \lbrack \mbox{d}x^2 , x^3 \rbrack = \lbrack \mbox{d}x^3 , x^2
     \rbrack = \mbox{d}x^1 ,  \quad
     \lbrack \mbox{d}x^3 , x^3 \rbrack = \mbox{d}x^2
  \ee
  and the left- (right-) partial derivatives are
  \be
     D_{\pm 1} = \partial_1 \pm \partial_2 \partial_3
     + {1 \over 6} \, \partial_3^3, \quad
     D_{\pm 2} = \partial_2 \pm {1 \over 2} \,
     \partial_3^2 , \quad
     D_{\pm 3} = \partial_3  \; .
  \ee
  The differential of a function $f$ thus involves third order
  derivatives,
  \be
     \mbox{d} f = ( \partial_1 + \partial_2 \partial_3
     + {1 \over 6} \, \partial_3^3 )f \; \mbox{d}x^1
     + ( \partial_2 + {1 \over 2} \, \partial_3^2 )f \; \mbox{d}x^2
     + \partial_3 f \; \mbox{d}x^3  \; .
  \ee
  For the associated algebra the index is 3.
  A generalization of this new calculus to $n$ dimensions with
  up to $n$th order left-partial derivatives will be described
  in the next subsection.
\item In this case we have
  \be
     \lbrack \mbox{d}x^1 , x^3 \rbrack = \lbrack \mbox{d}x^3 , x^1
     \rbrack = \mbox{d}x^1 ,  \quad
     \lbrack \mbox{d}x^2 , x^3 \rbrack = \lbrack \mbox{d}x^3 , x^2
     \rbrack = \mbox{d}x^2 ,  \quad
     \lbrack \mbox{d}x^3 , x^3 \rbrack = \mbox{d}x^3
  \ee
  with the left- (right-) partial derivatives
  \be
     D_{\pm 1} = \partial_1 \, \exp(\pm \partial_3), \quad
     D_{\pm 2} = \partial_2 \, \exp(\pm \partial_3), \quad
     D_{\pm 3} = \pm (\exp(\pm \partial_3) - 1) \; .
  \ee
\item Here we have the nonvanishing commutators
  \be
     \lbrack \mbox{d}x^1 , x^3 \rbrack = \lbrack \mbox{d}x^3 , x^1
     \rbrack = \mbox{d}x^1 ,  \quad
     \lbrack \mbox{d}x^2 , x^2 \rbrack = \mbox{d}x^2
                                                \nonumber      \\
     \lbrack \mbox{d}x^2 , x^3 \rbrack = \lbrack \mbox{d}x^3 , x^2
     \rbrack = \mbox{d}x^2 , \quad
     \lbrack \mbox{d}x^3 , x^3 \rbrack = \mbox{d}x^3
  \ee
  and the left- (right-) partial derivatives
  \be
     D_{\pm 1} &=& (\partial_1 \pm {1 \over 2} \partial_2^2)
     \, \exp(\pm \partial_3), \nonumber \\
     D_{\pm 2} &=& \partial_2 \, \exp(\pm \partial_3) \\
     D_{\pm 3} &=& \pm (\exp(\pm \partial_3) - 1) \; .
                                           \nonumber
  \ee
\end{enumerate}
                                     \hfill {\Large $\Box$}
\vskip.3cm
\noindent
Let us consider a differential calculus which is an extension in
the sense of subsection 4.1. $\check{D}_I, \, I=1, \ldots, n+1,$ are
the corresponding left-partial derivatives and $D_j, \, j=1,
\ldots, n,$ those of the $n$-dimensional calculus which generates the
extension. Then we have the following result.
\vskip.3cm
\noindent
{\bf Lemma 4.5}
\be
  \check{D}_j &=& D_j \, \exp(\partial_{n+1})
     \qquad  (j=1, \ldots, n)   \\
  \check{D}_{n+1} &=& \exp(\partial_{n+1}) - 1  \; .
\ee
\vskip.1cm
\noindent
{\bf Proof:} Recalling (\ref{df-inner}), we find
\begin{eqnarray*}
       \mbox{d} f
   &=& \lbrack dx^{n+1} , f \rbrack \\
   &=& \left. \left( \exp \left( \check{\bf C}^I \, {\partial \over
       \partial y^I} \right) - {\bf 1} \right)^{n+1}_J \, f(y)
       \right |_{y=x} \; \mbox{d} x^J
                         \qquad \mbox{using (\ref{dx^i-f-comm})} \\
   &=& \left. \left( \exp \left( \check{\bf C}^i \, {\partial \over
       \partial y^i} \right) \, \exp \left({\partial \over
       \partial y^{n+1}} \right) - {\bf 1} \right)^{n+1}_J \, f(y)
       \right |_{y=x} \; \mbox{d} x^J     \\
   &=& \check{D}_J f \; \mbox{d} x^J   \; .
\end{eqnarray*}
On functions which do not depend on $x^{n+1}$ the $\check{D}_j$
coincide with the operators $D_j$. Hence,
\be
  D_j = \left. \left( \exp \left( \check{\bf C}^i \, {\partial
               \over \partial y^i} \right) - {\bf 1} \right)^{n+1}_j
               \, f(y) \right |_{y=x}     \; .
\ee
With this observation, the conjectured formulas follow immediately.
                                            \hfill {\Large $\Box$}
\vskip.3cm
\noindent
In the last two examples -- {\bf (11)} and {\bf (12)} -- treated above
we have special cases of this general result.

\subsection{An $n$-dimensional differential calculus with up to
            $n$th order partial derivatives}
The relations
\be
   \lbrack \mbox{d}x^i , x^j \rbrack = \left \lbrace
       \begin{array}{cl}
       \mbox{d}x^{i+j} \quad & \mbox{if } i+j \leq n  \\
            0                & \mbox{otherwise}
       \end{array}  \right.
\ee
determine a consistent differential calculus on $\cal A$. For $n=2$
this is our calculus \framebox[1cm]{\bf I} and for $n=3$ we recover the
calculus {\bf (10)} of section 4.4 (up to a renumbering of the $x^i$).
\vskip.3cm

A partition $p(m)$ of a positive integer $m$ is a nonincreasing
sequence of positive integers $p_1, \ldots, p_r$ such that
$\sum_{s=1}^r p_s = m$. It is always possible to write $p(m)$ in the
form $(1^{k_1}, 2^{k_2},\ldots, m^{k_m})$ where $\ell^k$ means
that $\ell$ appears exactly $k$ times in $p(m)$. With the
definitions
\be
   p(m)! := (k_1 !) \cdots (k_m !), \quad
   \partial_{p(m)} := \partial_{p_1} \cdots \partial_{p_r}
\ee
one finds
\be
     \mbox{d}f = \sum_{m=1}^n \, \sum_{p(m)} {1 \over p(m)!} \,
                 \partial_{p(m)} f \; \mbox{d}x^m    \; .
\ee
The first four left-partial derivatives are thus
\be
   D_1 &=& \partial_1       \nonumber               \\
   D_2 &=& \partial_2 + {1 \over 2} \, \partial_1^2
                                   \nonumber               \\
   D_3 &=& \partial_3 + \partial_2 \partial_1
                  + {1 \over 6} \, \partial_1^3
                                   \nonumber               \\
   D_4 &=& \partial_4 + \partial_3 \partial_1
                  + {1 \over 2} \, \partial_2^2
                  + {1 \over 2} \, \partial_2 \partial_1^2
                  + {1 \over 24} \, \partial_1^4  \; .
\ee

\subsection{On some solutions of the consistency conditions}
We can always decompose the structure functions $C^{ij}{}_k$
as follows,
\be                            \label{C-decomp}
   C^{ij}{}_k = {1 \over n+1} \left ( \delta^i_k \, C^j
                + \delta^j_k \, C^i \right ) + P^{ij}{}_k
\ee
where
\be
     C^i := C^{ij}{}_j  \, , \quad P^{ij}{}_j = 0 \; .
\ee
The first of the consistency conditions (\ref{C-conds}) requires
that
\be
     P^{ij}{}_k = P^{ji}{}_k  \; ,
\ee
the second becomes
\be
   P^{ik}{}_m \, P^{jm}{}_\ell - P^{jk}{}_m \, P^{im}{}_\ell
   + {1 \over n+1} \, C^m \, \left( \delta^j_\ell \, P^{ik}{}_m
                - \delta^i_\ell \, P^{jk}{}_m \right )
           & &     \nonumber \\
   + {1 \over (n+1)^2} \, C^k \left ( \delta^j_\ell \, C^i
                - \delta^i_\ell \, C^j \right ) &=& 0  \; .
\ee
For vanishing $P^{ij}{}_k$ this implies $C^i = 0$ and thus
$C^{ij}{}_k = 0$. A noncommutative differential calculus therefore
has to have a nonvanishing traceless part of $C^{ij}{}_k$.
\vskip.3cm
\noindent
In the following, we consider some solutions of the consistency
conditions for general $n$. With minor modifications these are taken
from \cite{Boug+Fair86} where the consistency conditions (\ref{C-conds})
arose in a different context.
Instead of using (\ref{C-decomp}) it appears to be more
convenient to use a corresponding decomposition with $C^i$ replaced by
vector components $P^i$ and $P^{ij}{}_k$ not necessarily traceless.
A simple solution of (\ref{C-conds}) is then
\be
   C^{ij}{}_k =  \delta^i_k \, P^j + \delta^j_k \, P^i - P^i P^j U_k
\ee
with an additional covector\footnote{$P^i$ and $U_j$ transform as
the components of a vector and a covector, respectively, under
$GL(n,\Cx)$- or, more generally, $GL(n,{\cal A})$-transformations,
cf section 4.2.} $U$ subject to
\be                               \label{UP}
          U_k \, P^k = 1  \; .
\ee
A generalization of this solution is given by
\be
 C^{ij}{}_k = {1 \over \Delta} \, \left | \begin{array}{ccccc}
 \delta^i_k \, P^j_1 + \delta^j_k \, P^i_1 - P^i_1 P^j_1 U_k &
 \delta_{k \ell} P^\ell_1 &  \delta_{k \ell} P^\ell_2 & \ldots &
 \delta_{k \ell} P^\ell_L  \\
 0 & P_1 \cdot P_1 & P_1 \cdot P_2 & \ldots & P_1 \cdot P_L \\
 - (P^i_1 - P^i_2)(P^j_1 -P^j_2) & P_2 \cdot P & P_2 \cdot P_2 &
 \ldots & P_2 \cdot P_L \\
 \vdots & \vdots & \vdots & \cdots & \vdots \\
 - (P^i_1 - P^i_L)(P^j_1 -P^j_L) & P_L \cdot P & P_L \cdot P_1 &
 \ldots & P_L \cdot P_L
 \end{array} \right |           \label{C-det}
\ee
in terms of a determinant (cf \cite{Boug+Fair86}). Here $P_1, \ldots,
P_L$ are $L \leq n$ linearly independent vectors, all subjected to the
condition (\ref{UP}). Furthermore, we have introduced the abbreviation
$P \cdot P' := \sum_{i = 1}^n P^i \, {P'}^i$ and the subdeterminant
\be
 \Delta := \left | \begin{array}{cccc}
   P_1 \cdot P_1 & P_1 \cdot P_2 & \ldots & P_1 \cdot P_L   \\
   P_2 \cdot P_1 & P_2 \cdot P_2 & \ldots & P_2 \cdot P_L   \\
   \vdots & \vdots & \cdots & \vdots \\
   P_L \cdot P & P_L \cdot P_2 & \ldots & P_L \cdot P_L
                   \end{array} \right |   \; .
\ee
The expression (\ref{C-det}) is obviously symmetric in the two
indices $i,j$. It satisfies $C^{ij}{}_k P^k_\alpha = P^i_\alpha
P^j_\alpha, \, \alpha = 1, \ldots, L,$ which can then be used to
prove that also the second condition in (\ref{C-conds}) is satisfied
(cf \cite{Boug+Fair86}).\footnote{We were unable to verify the
statement in \cite{Boug+Fair86} that certain linear combinations of
terms of the form (\ref{C-det}) (cf (28) in \cite{Boug+Fair86}) also
satisfy the nonlinear condition in (\ref{C-conds}). A counter example
is given by $n=3, L=2$ with $P^i_1 = \delta^i_1,
P^i_2 = \delta^i_2$ (and $U_k = \delta^1_k + \delta^2_k$).}
A different proof is given below which moreover provides us with a
clear characterization of the differential calculi determined by
(\ref{C-det}).
\vskip.3cm
\noindent
{\bf Lemma 4.6} Via a $GL(n,{\cal A})$-transformation, (\ref{C-det})
is equivalent to
\be
   C^{ij}{}_k = \sum_{\alpha=1}^{L-1} \delta^i_\alpha \delta^j_\alpha
                \delta^\alpha_k + \sum_{J=L}^{n-1} \delta^j_n
                \delta^i_J \delta^J_k
                + \delta^i_n \, \left(\delta^j_n \delta^n_k
                + \sum_{J=L}^{n-1} \delta^j_J \delta^J_k \right) \; .
\ee
{\bf Proof:} After a cyclic renumbering of the vectors $P_1, \ldots,
P_L$, (\ref{C-det}) takes the form
\begin{eqnarray*}
   C^{ij}{}_k = {1 \over \Delta} \, \left| \begin{array}{ccc}
   \delta^i_k \, P^j_n + \delta^j_k \, P^i_n - P^i_n P^j_n U_k &
   \delta_{k \ell} \, P^\ell_\beta  & \delta_{k \ell} \, P^\ell_n \\
   - (P^i_n-P^i_\alpha) (P^j_n-P^j_\alpha) & P_\alpha \cdot P_\beta
   & P_\alpha \cdot P_n \\
   0 & P_n \cdot P_\beta & P_n \cdot P_n
   \end{array} \right|
\end{eqnarray*}
where $\alpha, \beta = 1,\ldots,L-1$. Now we complete the set of
linearly independent vectors $P_\alpha, P_n$ to a linear frame
(field) by adding vectors $P_J, \, J=L, \ldots, n-1,$ such that
$P_\alpha \cdot P_J = P_n \cdot P_J = 0$. Then
\begin{eqnarray*}
 C^{ij}{}_k P^k_\alpha = P^i_\alpha P^j_\alpha \, , \quad
 C^{ij}_k P^k_n = P^i_n P^j_n  \, , \quad
 C^{ij}{}_k P^k_J = P^{(i}_n P^{j)}_J - P^i_n P^j_n \, V_J
\end{eqnarray*}
where $V_J := U_k P^k_J$. Let $P \in GL(n, {\cal A})$ be the matrix
with entries $P^i_j$. The transformation ${x'}^k =
(P^{-1})^k_\ell \, x^\ell$ preserves (\ref{dx-C}) if ${\bf C'}^i =
(P^{-1})^i_j \, (P^{-1} {\bf C}^j P)$. Here it leads to
\begin{eqnarray*}
 {C'}^{ij}{}_k = \sum_\alpha \delta^i_\alpha \delta^j_\alpha
                 \delta^\alpha_k
                 + \sum_J \delta^{(i}_n \delta^{j)}_J \delta^J_k
                 + \delta^i_n \delta^j_n \, \left( \delta^n_k -
                 \sum_J V_J \delta^J_k \right)
\end{eqnarray*}
so that
\begin{eqnarray*}
 {\bf C}^\alpha = \left( \begin{array}{c|c|c}
                  E^r & 0 & 0 \\
                  \hline 0  & 0 & 0 \\
                  \hline 0  & 0 & 0
                         \end{array}\right) , \quad
 {\bf C}^J = \left( \begin{array}{c|c|c} 0 & 0 & 0 \\
                               \hline 0  & 0 & 0 \\
                               \hline 0  & e^J & 0
                    \end{array} \right) , \quad
 {\bf C}^n=\left( \begin{array}{c|c|c}  0 & 0 & 0 \\
                               \hline 0  & I & 0 \\
                               \hline 0  & -V & 1
                  \end{array} \right) ,
\end{eqnarray*}
where $(E^\alpha)^\beta_\gamma = \delta^\beta_\alpha
\delta^\alpha_\gamma$ and $(e^J)_K = \delta^J_K$. $I$ is the
$(n-L)\times (n-L)$ unit matrix.
A further $GL(n,{\cal A})$-transformation with
\begin{eqnarray*}
    A = \left( \begin{array}{c|c|c}  I  & 0 & 0 \\
                               \hline 0  & I & 0 \\
                               \hline 0  & V & 1
               \end{array} \right)
\end{eqnarray*}
eliminates the $V$-term.    \hfill {\large $\Box$}
\vskip.3cm
\noindent
According to the Lemma, the calculus determined by (\ref{C-det})
is $\hat{\otimes}^n \fbox{\bf L}$ for $L=n$ and
$(\hat{\otimes}^{L-1}\fbox{\bf L}) \hat{\otimes} \mbox{\bf Ext}
   (\hat{\otimes}^{n-L} \fbox{\bf K})$ for $L<n$ where
$\mbox{\bf Ext}$ indicates an extension in the sense of section 4.1.
Comparison with the list of calculi in subsection 4.4 shows that
the ansatz (\ref{C-det}) by far does not exhaust the possibilities.

One can try other ans\"atze, but it is unlikely that this can be
made into a systematic procedure to obtain the complete set of
solutions of the consistency conditions (\ref{C-conds}).

\section{Conclusions}
\setcounter{equation}{0}
In this work we have started a systematic exploration of differential
calculi on commutative algebras and presented several new results.
\vskip.2cm

A central part of this work is the classification of
3-dimensional differential calculi with constant structure functions
(on a commutative algebra with three generators). Much of the additional
material in this paper provides the necessary background or arose from
insights obtained via this classification. Apart from having solved
the classification problem for $n \leq 3$, we have presented
generalizations to arbitrary dimension $n$ for all the calculi
found in this way.
\vskip.2cm

Only four of the 3-dimensional calculi (more precisely:
$GL(3,\Cx)$-orbits) obtained in section 4.4 turned out to be
irreducible, i.e., they are not skew tensor products of
lower-dimensional calculi. Two of these are extensions of 2-dimensional
calculi (in the sense of section 4.1). They are `nonlocal' in the sense
that their left- (or right-) partial derivatives involve finite
difference operators. The remaining two genuinely 3-dimensional
calculi have local (though higher order) left- and right-partial
derivatives.
For one of them, a generalization to arbitrary dimension is already
known \cite{DMH92-grav,DMH93-stoch}. A corresponding generalization
of the other calculus is presented in section 4.7, the left-partial
derivatives are differential operators of up to $n$th order.
\vskip.2cm

Our classification procedure extends to $n > 3$, but the corresponding
calculations become much more involved. Computer algebra should
then be helpful.
\vskip.2cm

For the new calculi we were so far unable to find a relation
with structures of interest in other branches of mathematics or in
physics, similar to what we have for the examples mentioned in the
introduction. Further investigation of these calculi is
therefore required.
There is, however, a general aspect which supports our investigation
from a physical point of view. A study of differential calculi on
finite sets has shown that the choice of a differential calculus
assigns to a set a structure which should be regarded as an analogue
of that of a differentiable manifold \cite{DMH94-discr}.
Our present paper provides new examples of such generalized spaces
which can be regarded as deformations of $\Rl^n$ with the
ordinary differential calculus. Such a deformation induces (in a
universal manner) corresponding deformations of models and theories
built on the differential calculus (cf
\cite{DMHS93,DMH92-grav,DMH94-discr}). There is the hope to obtain
in this systematic way physical models which are somehow close to
known models but which improve the latter by properties like
complete integrability or finiteness (of quantum perturbation
theory, in particular for nonrenormalizable theories like gravity).

\begin{appendix}
\renewcommand{\theequation} {\Alph{section}.\arabic{equation}}

\section{On generalized differential calculi}
\setcounter{equation}{0}
Let $\cal A$ be a commutative associative algebra generated by
$x^i$ ($i=1,\ldots,n$), $\Theta$ an $\cal A$-bimodule which is free
of rank $m$ as a left $\cal A$-module, and $\mbox{d} \, : \, {\cal A}
\rightarrow \Theta$ a derivation. Here we do not assume that
the $\mbox{d}x^i$ are a left $\cal A$-module basis (as we did in
section 4). Moreover we will not even demand
that $\mbox{d} {\cal A}$ generates $\Theta$ as an $\cal A$-bimodule.
In this sense we generalize the notion of a (first order) differential
calculus as defined in section 2.\footnote{Such a generalization is
encountered if one tries to extend the correspondence between
differential calculi on finite (or discrete) sets and special digraphs
(see \cite{DMH94-discr}) to general digraphs (with multiple arrows
and loops). This is of interest for the study of processes on (e.g.,
electrical) networks and will be discussed elsewhere.}
\vskip.3cm

Let $\theta^\mu, \, \mu=1,\ldots,m,$ be a left $\cal A$-module basis
of $\Theta$. Every element $\varphi \in \Theta$ can then be expressed
as $\varphi = \varphi_\mu \, \theta^\mu$ with $\varphi_\mu \in {\cal
A}$. In particular, $\mbox{d}x^i = \gamma^i_\mu \, \theta^\mu$ where
$\gamma$ is an $n\times m$-matrix with entries in $\cal A$. More
generally,
\be
   \mbox{d}f = (\nabla_\mu f) \, \theta^\mu  \qquad (\forall f \in
                                                     {\cal A})
\ee
which defines linear operators $\nabla_\mu \, : \, {\cal A} \rightarrow
{\cal A}$. As a further consequence of the assumption that $\Theta$
is free as a left $\cal A$-module we have
\be                                    \label{theta-C}
   \lbrack \theta^\mu , x^i \rbrack =  C^{i\mu}{}_\nu \, \theta^\nu
\ee
with structure functions $C^{i\mu}{}_\nu \in {\cal A}$. The latter
constitute a set of $m \times m$ matrices ${\bf C}^i$. These
have to satisfy the following consistency conditions which are
derived in the same way as those in section 4,
\be
 \gamma^i_\mu \, C^{j \mu}{}_\nu & = & \gamma^j_\mu \, C^{i\mu}{}_\nu
 \qquad \quad
 (\forall \ i,j=1,\dots,n \quad \forall \mu = 1,\dots,m) ,
                                                     \nonumber  \\
 C^{i\mu}{}_\lambda \, C^{j \lambda}{}_\nu & = &
 C^{j\mu}{}_\lambda \, C^{i\lambda}{}_\nu
 \qquad (\forall \ i,j=1,\dots,n \quad \forall \mu , \nu = 1,\dots,m)
  \, .               \label{ext-consist}
\ee
The following is a generalization of Lemma 4.1.
\vskip.3cm
\noindent
{\bf Lemma A.1} The derivation $\mbox{d}$ is inner if and only if there
are $\rho_\mu \in {\cal A}$ ($\mu = 1, \ldots, m$) such that
$\rho_\mu \, C^{i \mu}{}_\nu = \gamma^i_\nu$. Then
$\mbox{d} f = \lbrack \rho_\mu \, \theta^\mu , f \rbrack$.
\vskip.1cm \noindent
{\bf Proof:} \\
``$\Rightarrow$'': follows from (\ref{theta-C}) and
  $\mbox{d}x^i = \lbrack \rho , x^i \rbrack$. \\
``$\Leftarrow$'': Define a derivation $\mbox{d}'$ by $\mbox{d}'f
  := \lbrack \rho , f \rbrack$. Then
\begin{eqnarray*}
 \mbox{d}'x^i = \lbrack \rho , x^i \rbrack = \rho_\mu \, \lbrack
 \theta^\mu , x^i \rbrack = \rho_\mu \, C^{i\mu}{}{}_\nu \,
 \theta^\nu = \gamma^i_\nu \, \theta^\nu = \mbox{d}x^i
\end{eqnarray*}
shows that $\mbox{d}'$ and $\mbox{d}$ coincide on the generators of
$\cal A$. Since both are derivations, we have $\mbox{d}' = \mbox{d}$.
                                   \hfill    {\Large $\Box$}
\vskip.3cm
The operators $\nabla_\mu$ are not derivations, in general. If $\cal
A$ is freely generated by the $x^i$, they satisfy the following twisted
Leibniz rule,
\be
   \nabla_\mu (fg) = \left. (\nabla_\nu f) \;
   \left( \exp \left( {\bf C}^i(x) \frac{\partial}{\partial y^i} \right)
   \right)^\nu_\mu g(y) \right|_{y=x} + f \, \nabla_\mu g
\ee
(where $y^i$ are auxiliary variables, cf section 4.5). We have the
following generalization of Lemma 4.3.
\vskip.3cm
\noindent
{\bf Lemma A.2} Let $\cal A$ be freely generated by the $x^i$.
In terms of ordinary partial derivatives the linear
operators $\nabla_\mu$ are then given by
\be                           \label{nabla-partial}
  \nabla_\mu = \gamma^i_\mu \partial_i + \sum_{r=2}^\infty
               \frac{1}{r !} \, \gamma^{k_1}_\nu ({\bf C}^{k_2}
               \cdots {\bf C}^{k_r} )^\nu{}_\mu \,
               \partial_{k_1} \dots \partial_{k_r} \; .
\ee
\hfill    {\Large $\Box$}
\vskip.3cm
\noindent
We omit the proof which is a minor modification of the proof of
Lemma 4.3.

\vskip.3cm

Using the construction of section 2.4 one obtains an extension
$\check{\Theta}$ of the $\cal A$-bimodule $\Theta$ such that
\be
  \lbrack \theta^\mu , x^i \rbrack
               = C^{i\mu}{}_\nu \, \theta^\nu , \quad
  \lbrack \theta^{m+1} , x^i \rbrack
               = \mbox{d} x^i = \gamma_\mu^i \, \theta^\mu
\ee
where $\theta^{m+1}$ corresponds to $\rho$ in section 2.4.
The two relations can be combined to
\be
   [\theta^{\check{\mu}},x^i] = \check{C}^{i\check{\mu}}{}_{\check{\nu}}
                              \, \theta^{\check{\nu}}
\ee
where the $\check{C}^{i\check{\mu}}{}_{\check{\nu}}$ are the entries of
the $(m+1)\times (m+1)$ matrices
\be
   {\bf \check{C}}^i : = \left( \begin{array}{c|c}
                       {\bf C}^i & \begin{array}{c} 0\\ \vdots \\ 0\\
                                   \end{array} \\ \hline
                       \gamma^i_1 \dots \gamma^i_m & 0\\
                               \end{array}   \right)  \; .
\ee
These matrices indeed satisfy the consistency conditions
(\ref{ext-consist}) (if the matrices ${\bf C}^i$ satisfy them).

\end{appendix}

\end{document}